\pdfoutput=1

\documentclass[11pt]{article}

\usepackage[final]{acl}

\usepackage{times}
\usepackage{latexsym}

\usepackage[T1]{fontenc}

\usepackage[utf8]{inputenc}

\usepackage{microtype}

\usepackage{inconsolata}

\usepackage{url}            
\usepackage{booktabs}       
\usepackage{amsfonts}       
\usepackage{nicefrac}       
\usepackage{microtype}      
\usepackage{xcolor}         
\usepackage{enumitem}
\usepackage{bbding}
\usepackage{array}
\newcolumntype{L}[1]{>{\raggedright\let\newline\\\arraybackslash\hspace{0pt}}m{#1}}
\newcolumntype{C}[1]{>{\centering\let\newline  \\\arraybackslash\hspace{0pt}}m{#1}}
\newcolumntype{R}[1]{>{\raggedleft\let\newline \\\arraybackslash\hspace{0pt}}m{#1}}
\usepackage{graphicx}
\usepackage{subcaption}
\usepackage{url}
\usepackage{bm}
\usepackage{amsmath,amssymb,amsfonts,amsthm}
\usepackage{caption} 
\usepackage[normalem]{ulem}
\usepackage{float} 
\usepackage{verbatim}
\usepackage{algorithm}
\usepackage{algorithmic}
\usepackage{multirow}
\usepackage{wrapfig}
\usepackage{ulem}
\usepackage[most]{tcolorbox}
\usepackage{makecell}

\title{MAS-on-the-Fly: In-Context Structural Adaptation of LLM-Based Multi-Agent Systems}

\author{%
	Guangyi Liu$^{1}$, Haojun Lin$^{2}$, Huan Zeng$^{2}$, Heng Wang$^2$, Quanming Yao$^{1,3,4}$
		\thanks{Correspondence is to Q.Yao.} 
	\\
	$^1$Department of Electronic Engineering, Tsinghua University \\ 
	$^2$Ant Group\\
	$^3$State Key Laboratory of Space Network and Communications \\
	$^4$Beijing National Research Center for Information Science and Technology \\
	\texttt{liugy24@mails.tsinghua.edu.cn, qyaoaa@tsinghua.edu.cn} 
}

\begin{document}
\maketitle

\begin{abstract}
Large Language Model (LLM)-based multi-agent systems (MAS) have emerged as a promising paradigm for complex tasks.
However, existing works often rely on manual designs or ``one-size-fits-all'' automation and lack adaptability after deployment.
We study \emph{in-context structural adaptation}, where structured experience conditions both query-dependent system generation and execution-time reconfiguration without updating LLM parameters.
We introduce MASFly, which realizes this adaptation through two complementary mechanisms.
First, a retrieval-augmented SOP instantiation mechanism retrieves and adapts successful collaboration patterns to construct a query-specific MAS.
Second, an experience-enhanced process supervision mechanism uses a dedicated Watcher agent to monitor execution against prior failure experience and reconfigure the system upon abnormal behavior.
Experiments demonstrate that MASFly achieves state-of-the-art performance, including a 61.7\% success rate on TravelPlanner, with strong task adaptability and robustness.
Our code and data are available at \url{https://github.com/LARS-research/MASFly}.
\end{abstract}

\section{Introduction}

Large Language Models (LLMs) have shown remarkable reasoning and generation abilities~\cite{achiam2023gpt,liu2024deepseek, yang2025qwen3}.
Yet complex, multi-faceted problems often exceed a single LLM's capacity~\cite{talebirad2023multi}.
This has motivated LLM-based multi-agent systems (MAS)~\cite{hong2023metagpt}, in which agents assume specialized roles and collaborate on advanced problem-solving.

\begin{figure}[ht]
	\centering
	\vspace{-2px}
	\includegraphics[width=0.48\textwidth]{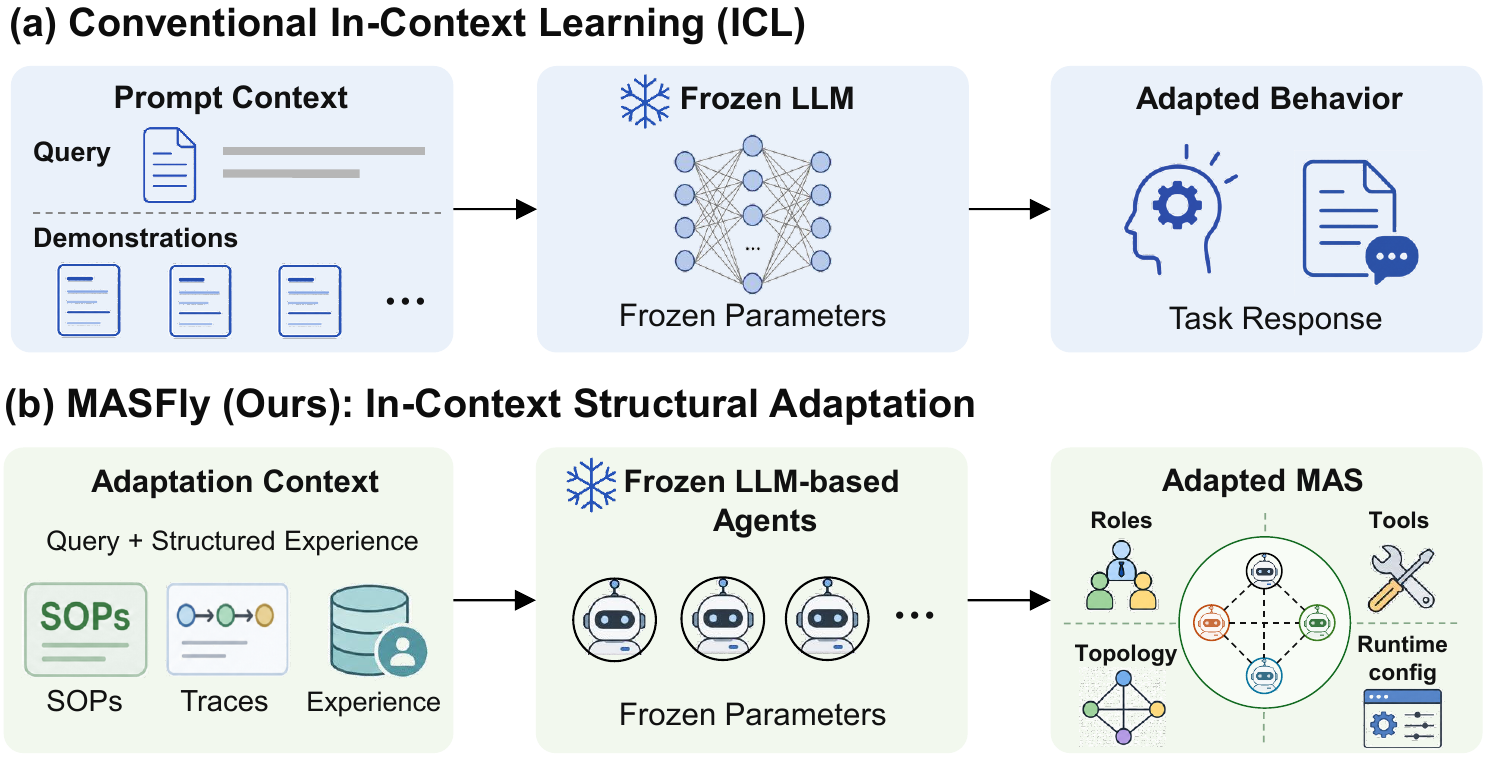}
	\vspace{-18px}
	\caption{Analogy between conventional in-context learning (ICL) and our in-context structural adaptation. ICL conditions a frozen model on demonstrations to adapt model behavior; MASFly conditions frozen LLM agents on structured collaboration experience to adapt the multi-agent system's roles, tools, communication topology, and execution-time configuration.}
	\label{fig:motiv}
	\vspace{-20px}
\end{figure}

Early LLM-based MASs, such as MetaGPT~\cite{hong2023metagpt} and ChatDev~\cite{qian2023chatdev},
construct systems with handcrafted standard operating procedures (SOPs).
Later work automates individual components, including agent-role generation (e.g., AgentVerse~\cite{chen2023agentverse}) and communication-topology learning with external modules (e.g., GPTSwarm~\cite{zhuge2024gptswarm}).
Recent studies automate full MAS construction by prompting LLMs to search a predefined design space for a single ``optimal'' system (e.g., AgentSquare~\cite{shang2024agentsquare}), or by training a meta-agent to generate systems for different queries (e.g., MAS-GPT~\cite{ye2025mas}).
Yet adapting the system itself in context through query-dependent system generation and execution-time reconfiguration remains underexplored.
Existing search-based methods rely on one-size-fits-all construction that cannot adapt to query-specific needs, while training-based methods require substantial data and training before testing. 
Moreover, most prior works neglect execution-time adaptation, leading to brittle, open-loop pipelines that struggle to recover from agent failures or unexpected environmental feedback. 
This motivates a central question:
\textit{Can an LLM-based MAS adapt its structure in context---conditioning on prior collaboration experience and the evolving execution trajectory---without updating model parameters?}

We refer to this capability as \emph{in-context structural adaptation}: conditioning on structured experience to instantiate or reconfigure a MAS's composition, agent specifications, tool assignments, and communication topology without updating model parameters.
Whereas conventional in-context learning (ICL) adapts the behavior of a fixed model through contextual demonstrations~\cite{wu2025why}, our setting extends this conditioning principle to the organization and execution of an entire multi-agent system.
We subsequently use \emph{in-context adaptation}.

To address this problem, we propose MASFly, a multi-agent framework that extends ICL's conditioning principle from model behavior to system structure, as illustrated in Figure~\ref{fig:motiv}.
For adaptive system generation, MASFly retrieves successful collaboration patterns from an SOP repository and instantiates a query-specific MAS.
For the non-stationary execution environment, an experience-enhanced process supervision mechanism uses a global Watcher agent and personalized experience pool to monitor execution and enable on-the-fly reconfiguration (e.g., replacing malfunctioning agents), yielding a resilient closed-loop process.
Extensive experiments show that MASFly achieves state-of-the-art performance, including a 61.7\% success rate on TravelPlanner.
The contributions are summarized as follows:
\begin{itemize}[leftmargin=*,itemsep=2pt,topsep=2pt,parsep=0pt]
\item
We formulate in-context structural adaptation for LLM-based MAS, encompassing query-dependent system generation and execution-time reconfiguration without parameter updates.
\item
We propose MASFly with a retrieval-augmented SOP instantiation mechanism that reuses successful collaboration patterns to construct query-specific agent roles, tool assignments, and communication structures.
\item
We design an experience-enhanced process supervision mechanism,
which integrates a global Watcher agent and a personalized experience pool for execution-time reconfiguration.
\item
Experiments across diverse tasks show that MASFly outperforms strong baselines while exhibiting task adaptability and robustness.
\end{itemize}

\section{Related Work}

\begin{figure*}[ht]
	\centering
	\includegraphics[width=0.98\textwidth]{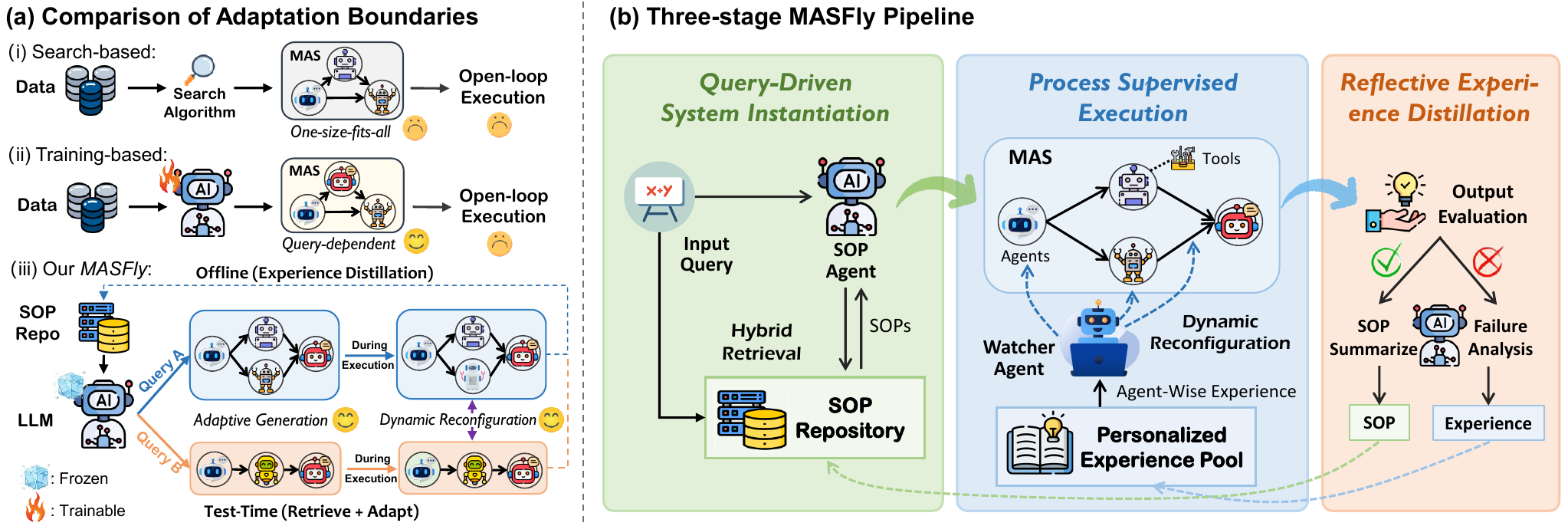}
	\vspace{-2px}
	\caption{Positioning and operational overview of MASFly. (a) Unlike search-based methods that deploy a one-size-fits-all system and training-based methods that learn a generator, MASFly retrieves structured experience to adapt a frozen LLM-based MAS at test time, including during execution. (b) MASFly instantiates a query-specific system, executes it under Watcher supervision, and distills successful SOPs and failure experience for future tasks.}
	\label{fig:framework}
	\vspace{-8px}
\end{figure*}

\subsection{LLM-based Multi-Agent System}
The evolution of LLM-based MAS has undergone a  transition from manually designed frameworks to increasingly automated and adaptive construction.
Early systems, such as MetaGPT~\cite{hong2023metagpt} and ChatDev~\cite{qian2023chatdev}, rely heavily on handcrafted SOPs to configure agent roles and workflows.
Subsequent research introduces semi-automated mechanisms.
For example, AgentVerse~\cite{chen2023agentverse}, EvoAgent~\cite{yuan2024evoagent}, and MegaAgent~\cite{wang2025megaagent} leverage LLMs to automatically generate diverse agent roles.
Other studies, such as GPTSwarm~\cite{zhuge2024gptswarm} and G-Designer~\cite{zhang2024g}, employ external models to learn the communication topology across agents.

Recently, the focus has shifted toward fully automated MAS generation driven by LLMs.
These methods typically prompt the LLM to generate executable code that represents the entire system.
\textit{Search-based methods} (e.g., ADAS~\cite{hu2024automated}, AFlow~\cite{zhang2024aflow}, AgentSquare~\cite{shang2024agentsquare}, SwarmAgentic~\cite{zhang2025swarmagentic} use LLMs to identify an "optimal" system configuration within a predefined search space.
However, these methods rely on fixed systems, lacking dynamic adaptability to diverse queries.
\textit{Training-based approaches}, such as MAS-GPT~\cite{ye2025mas} and FlowReasoner~\cite{gao2025flowreasoner}, train a meta-agent via supervised fine-tuning (SFT) or reinforcement learning (RL) to autonomously generate MAS configurations.
While more adaptive to different queries, these approaches need extensive data and struggle to generalize.
Moreover, most existing works neglect execution-time adaptability, leading to brittle open-loop pipelines.

\subsection{In-Context Learning and Test-Time Adaptation}

In-context learning (ICL) enables a frozen LLM to adapt its behavior by conditioning on demonstrations supplied in its context~\cite{dong2024survey,wu2025why}.
Retrieval-augmented generation (RAG)~\cite{lewis2020retrieval} retrieves relevant documents or memories at inference time to fill knowledge gaps and guide generation, while test-time scaling~\cite{snell2024scaling} allocates additional inference computation through sampling or reasoning-path exploration.
Test-time adaptation methods~\cite{wang2020tent}, by contrast, often adapt model parameters to unlabeled target-domain data.
These paradigms primarily adapt the knowledge, behavior, or computation of an individual model.
We extend the conditioning principle of ICL to multi-agent organization: retrieved structured experience guides how a MAS is instantiated and reconfigured, while the underlying LLM parameters remain fixed.
More broadly, data-efficient agentic learning studies how agents can acquire and reuse experience without relying solely on model scaling~\cite{wang2026beyond}.
MASFly contributes a complementary system-level perspective by supplying structured collaboration experience as context for system generation and reconfiguration.

\subsection{Memory in Multi-Agent System}

Memory plays a vital role in enabling LLM agents to accumulate experiences and enhance reasoning capabilities.
Research on single-agent memory has advanced considerably, 
evolving from simple RAG-style context extensions~\cite{packer2023memgpt} to cognitively inspired architectures that support cross-trial learning~\cite{xu2025mem,wang2024agent}.
However, memory mechanisms tailored for MAS remain largely underexplored~\cite{tang2025agent}.
Recent efforts (e.g., G-Memory~\cite{zhang2025g}) have attempted to manage MAS interaction histories and experiences via hierarchical graph structures, 
offering external memory banks for different MASs.

\section{Problem Description}

We study \emph{in-context structural adaptation} for LLM-based MAS.
At execution step $t$, we represent a system configuration as
\begin{equation}
	\mathcal{M}^{(t)} = \left(\mathcal{A}^{(t)}, \Pi^{(t)}, \mathcal{T}^{(t)}, \mathcal{G}^{(t)}\right),
\end{equation}
where $\mathcal{A}^{(t)}$ is the agent composition, $\Pi^{(t)}$ contains agent roles and instructions, $\mathcal{T}^{(t)}$ denotes tool assignments, and $\mathcal{G}^{(t)}$ is the communication topology.
Given a query, the goal is not to run a fixed, one-size-fits-all configuration, but to instantiate and adjust these components according to the current task and its evolving execution state.
Rather than updating model parameters, the system adapts by conditioning on task-relevant context and execution history.
Formally, for the query $q$, the model first constructs an initial query-specific system
\begin{equation}
	\mathcal{M}^{(0)} = \mathcal{I}_\theta(q \mid c_q),
\end{equation}
where $c_q$ denotes the task-relevant context used for initialization.
During execution, the system may further adjust its current configuration according to the intermediate execution state $h^{(t)}$ and additional contextual signals $c_h$:
\begin{equation}
	\mathcal{M}^{(t+1)} \leftarrow \mathcal{C}_\theta\!\left(\mathcal{M}^{(t)}, h^{(t)} \mid c_h\right).
\end{equation}
After a sequence of such adjustments, the final output is produced by the adapted system:
\begin{equation}
	y = \mathcal{E}_\theta(q, \mathcal{M}^{(T)}).
\end{equation}
During held-out evaluation, the LLM parameters and the experience repositories supplying $c_q$ and $c_h$ remain fixed.
Adaptation therefore occurs through contextual inference over the system configuration rather than through parameter or repository updates.


Achieving such adaptive system generation and execution is nontrivial.
First, constructing a suitable and high-quality MAS solely
using a pretrained LLM is challenging, as the design space (including agent roles, tools and communication topology) is vast and models may lack task-specific priors.
Second, even a strong initial design may fail during long-horizon execution, where errors can accumulate through miscoordination, tool misuse, or incorrect intermediate decisions, yet the agents themselves may lack awareness of these failures and thus be unable to correct them during execution.
These challenges call for a unified framework that supports both query-dependent system construction and execution-time correction through context.

\section{Proposed Method}

\subsection{Overall Framework}

To address the above challenges, MASFly adopts a closed-loop framework with three stages, as illustrated in Figure~\ref{fig:framework}:
(1) \emph{Query-Driven System Instantiation},
(2) \emph{Process-Supervised Execution}, and
(3) \emph{Reflective Experience Distillation}.
MASFly first performs dynamic system generation through a retrieval-augmented SOP instantiation mechanism, constructing a query-dependent MAS by conditioning on retrieved collaboration patterns from the SOP repository.
During execution, MASFly applies an experience-enhanced process supervision mechanism.
A global Watcher agent acts as a regulator, periodically monitoring the system with reference to a personalized experience pool and dynamically reconfiguring agents when abnormal behaviors are detected.
After execution, the LLM reflects on the outcome, distilling reusable experience to support future tasks.
Through this design, MASFly adapts its configuration to the current query and execution trajectory, while offline reflective distillation builds structured experience that supports subsequent held-out tasks.

\subsection{Query-Driven System Instantiation}
Different queries require distinct agent roles, tool allocations, and communication structures, rendering a fixed MAS design suboptimal across diverse tasks. 
However, generating adaptive systems on the fly for LLM is challenging due to the combinatorially large design space of MAS. 
To overcome this, we design a \textit{retrieval-augmented SOP instantiation mechanism} that retrieves historically relevant successful collaboration patterns and synthesizes them into a query-specific system.

\begin{figure}[ht]
	\centering
	\vspace{-2px}
	\includegraphics[width=0.87\columnwidth]{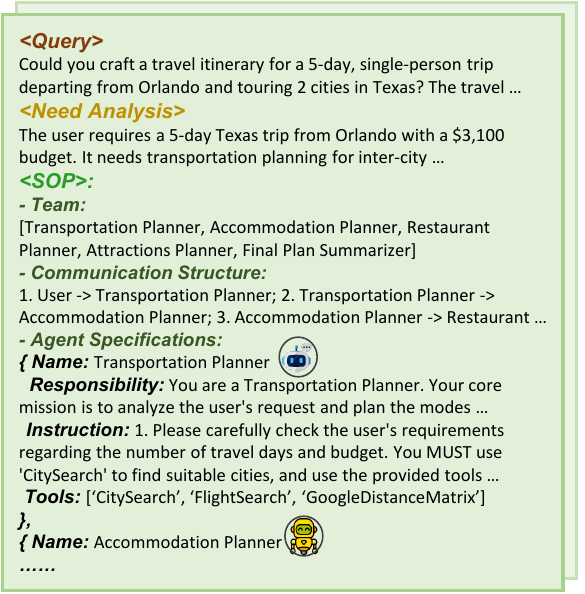}
	\vspace{-4px}
	\caption{The case illustration of the SOP Repository.}
	\vspace{-4px}
	\label{fig:SOPrepo}
\end{figure}

\subsubsection*{SOP Repository}

To capture successful collaboration experience, we use LLMs to automatically construct an SOP repository across diverse tasks (detailed in Section~\ref{ssec:reflect}). 
Each SOP, representing a reusable collaboration template, is associated with a case, which is formalized as a tuple \(c_i=(q_i,n_i,s_i)\).
Here, \(q_i\) is the \textit{User Query}; \(n_i\) is the LLM-generated \textit{Need Analysis}, detailing the task objective and required capabilities (e.g., specific tools or agent roles); and \textit{SOP} \(s_i\) is the multi-agent collaboration pattern. 
As shown in Figure~\ref{fig:SOPrepo}, each SOP records the \textit{team composition}, the \textit{communication structure} among agents (a directed graph with textual descriptions dictating information flow), and precise \textit{agent specifications} (including responsibilities, instructions, and accessible tools).
By systematically recording structured operational knowledge, the SOP repository enables the system to flexibly reuse and recombine proven collaborative strategies for new tasks.

\subsubsection*{Hybrid Retrieval and Instantiation}

Given a new query \(q\), MASFly first prompts an LLM to infer the corresponding need analysis \(n\), and then retrieves relevant cases from the SOP repository using a hybrid score that combines external query similarity and internal need similarity:
\begin{align}
	s(q, c_i)
	=
	\; & \lambda \cdot \mathrm{Sim}(q, q_i)
	\notag
	\\
	& + (1-\lambda) \cdot \mathrm{Sim}(n, n_i),
	\label{eq:retrieve}
\end{align}
where \(\lambda \in [0,1]\) balances the two terms, and \(\mathrm{Sim}(\cdot,\cdot)\) denotes cosine similarity computed from embeddings.
Then, the SOPs from the top-\(K\) retrieved cases guide the LLM to instantiate a query-specific operation procedure (OP):
\begin{equation}
	OP = \mathrm{LLM}(q, n, \{s_i\}_{i=1}^{K}),
\end{equation}
where \(OP\), structured similarly to an SOP, specifies the query-specific agent prompts and communication workflow, thereby defining the initial system \(\mathcal{M}^{(0)}\). 
This way, MASFly transforms reusable procedural templates into concrete MASs, preserving effective coordination patterns while flexibly adapting to queries.
Notably, MASFly can also operate in a zero-shot setting without repository support, relying solely on the LLM for orchestration.

\subsection{Process-Supervised Execution}

Even with a query-specific operation procedure, multi-agent execution remains inherently non-stationary: agents may deviate from the prescribed workflow, misuse tools, or encounter unexpected environmental feedback. 
To enable execution-time adaptability, we design an \textit{experience-enhanced process supervision mechanism} centered on a global Watcher agent, which monitors the running system and intervenes with reference to a Personalized Experience Pool (PEP) when abnormal behaviors are detected, thereby enabling adaptive and reliable execution.

\subsubsection*{Supervision and Intervention}

During system execution, supervision is triggered periodically.
A global Watcher agent examines the current execution context against the instantiated \(OP\) to detect abnormal behaviors. 
Specifically, the Watcher conducts dual-level supervision:
(1) \textit{Inter-Agent Communication}: verifying whether agents follow the prescribed workflow in the \(OP\), maintain output consistency, and avoid unproductive conversational loops;
(2) \textit{Agent--Environment Interaction}: assessing whether agents properly invoke designated tools, respond appropriately to dynamic contexts, and avoid repetitive or irrelevant outputs.

When an anomaly is detected, the Watcher intervenes and can take one of two actions:
(1) \textit{Experience-Guided Feedback}: sending corrective feedback to the faulty agent by incorporating insights from the PEP to improve its reasoning or operational strategy in real time;
(2) \textit{Dynamic Agent Replacement}: removing the critically flawed agent and dynamically instantiating a new one to resume the task, thereby reconfiguring the system and ensuring workflow continuity.

To balance controllability and efficiency, MASFly supports flexible supervision policies, such as triggering supervision every \(M\) communication rounds, every \(L\) environment-interaction steps, or under a budget on the total number of interventions. Related experiments are provided in Appendix~\ref{appendix:watcher}.

\begin{figure}[ht]
	\centering
	\vspace{-2px}
	\includegraphics[width=0.87\columnwidth]{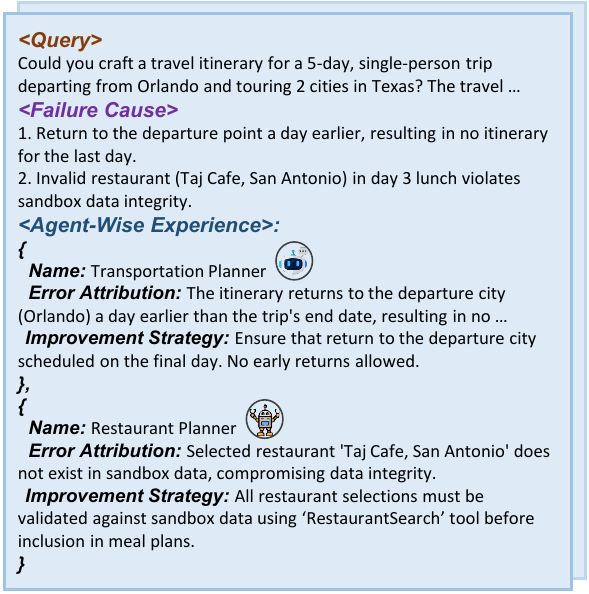}
	\vspace{-4px}
	\caption{The case illustration of the PEP.}
	\vspace{-4px}
	\label{fig:Personrepo}
\end{figure}

\subsubsection*{Personalized Experience Pool}

High-quality supervision requires the ability to identify and handle recurring anomalies. 
To this end, we further construct a Personalized Experience Pool (PEP) through LLM-based reflection (detailed in Section~\ref{ssec:reflect}), which stores common error patterns and their corresponding remedies to support precise intervention.
As shown in Figure~\ref{fig:Personrepo}, each record is formalized as \(r_i=(q_i, f_i, p_i)\), where \(q_i\) is the user query, \(f_i\) is the failure cause (e.g., task misallocation or tool misuse), and \(p_i=\{p_{i1},\dots,p_{im}\}\) is a set of agent-wise experiences. 
Each agent-wise experience contains an \textit{error attribution}, which identifies the agent’s contribution to the failure, and an \textit{improvement strategy}, which provides corrective guidance to prevent similar errors in future executions. 
Analogous to the retrieval over the SOP repository at initialization, relevant PEP records are retrieved once at the beginning and then used by the Watcher as experience priors for detecting likely pitfalls and selecting appropriate interventions throughout execution.

Through this experience-enhanced process supervision mechanism, MASFly treats execution as a closed-loop reconfiguration process rather than a fixed rollout from the initial \(OP\), thereby improving robustness and reliability in non-stationary multi-agent environments.

\begin{table*}[t]
	\centering
	\caption{Comparison of different methods.}
	\vspace{-4px}
	\label{tab:comparison}
	\small
	\begin{tabular}{l|>{\centering\arraybackslash}m{6.7em}>{\centering\arraybackslash}m{6.7em}>{\centering\arraybackslash}m{6.7em}>{\centering\arraybackslash}m{6.7em}|>{\centering\arraybackslash}m{6.7em}}
		\toprule
		Method & Adaptive Prompts & Adaptive Topology & Adaptive System Generation & Adaptive Execution &  Cross-Task Memory \\
		\midrule
		MetaGPT &   $\times$ & $\times$ & $\times $ & $\times$ & $\times$ \\ 
		AgentVerse & \checkmark & $\times$ & $\times$ & \checkmark & $\times$ \\
		MegaAgent & \checkmark & $\times$ & $\times$ & $\times$ & $\times$ \\
		GPTSwarm &  $\times$ & $\times$ & $\times$ & $\times$ & $\times$ \\ 
		G-Designer &  $\times$ & \checkmark & $\times$ & $\times$ & $\times$ \\ 
		AgentSquare &  $\times$ & $\times$ & $\times$ & $\times$ & \checkmark \\ 
		\midrule
		MASFly & \checkmark & \checkmark & \checkmark & \checkmark & \checkmark\\ 
		\bottomrule
	\end{tabular}
	\vspace{-2px}
\end{table*}

\begin{table*}[ht]
	\centering
	\caption{Performance comparison on TravelPlanner. Task-level bootstrap confidence intervals for Final Pass Rate are reported in Appendix~\ref{appendix:tp_bootstrap}.}
	\vspace{-4px}
	\small
	\begin{tabular}{clcccccc}
		\toprule
		\multirow{2}{*}{\textbf{Mode}} &\multirow{2}{*}{\textbf{Method}} & \multirow{2}{*}{\textbf{Delivery Rate}} & \multicolumn{2}{c}{\textbf{Commonsense}} & \multicolumn{2}{c}{\textbf{Hard Constraint}} & \multirow{2}{*}{\textbf{Final Pass Rate}} \\
		\cmidrule(lr){4-5} \cmidrule(lr){6-7}
		&&  & \textbf{Micro} & \textbf{Macro} &\textbf{ Micro} &\textbf{Macro} &  \\
		\midrule
		
		\multirowcell{6}{\textit{Two-}\\\textit{Stage}}
		&ReAct & 90.6 & 60.4 & 7.2 & 24.0 & 16.6 & 5.6 \\
		&Reflecxion & 89.4 & 62.8 & 7.8 & 24.5 & 17.2 & 7.8 \\
		&MetaGPT & 95.6 & 65.3 & 14.4 & 16.8 & 16.1 & 8.9 \\
		&AgentVerse & \textbf{100.0} & 66.7 & 4.8 & 15.2 & 10.6 & 2.2 \\
		&AgentSquare & 98.3 & 81.2 & 33.3 & 30.5 & 25.8 & 19.4 \\ 
		&G-Memory  & 96.7 & 84.4 & 36.1 & 37.1 & 29.4 & 24.4 \\
		\cmidrule(lr){2-8}
		&\textbf{MASFly} & 95.6 & \textbf{85.3} & \textbf{47.8} & \textbf{49.5} & \textbf{44.4} & \textbf{39.4} \\
		\midrule
		
		\multirowcell{7}{\textit{Sole-}\\\textit{Planning}} 
		&CoT & \textbf{100.0} & 75.8 & 13.3 & 55.2 & 48.9 & 6.7 \\
		&MetaGPT & 98.3 & 76.4 & 15.6 & 60.5 & 50.5 & 8.3 \\
		&EvoAgent & 95.6 & 69.2 & 13.9 & 59.5 & 47.8 & 8.3 \\
		&AgentVerse & \textbf{100.0} & 79.2 & 11.1 & 59.0 & 53.3 & 6.7 \\
		&MegaAgent & 94.4 & 82.9& 26.2& 60.5& 46.9& 10.0\\
		&SwarmAgentic & \textbf{100.0} & 88.0 & 53.6 & 62.4 & 50.9 & 31.1 \\\cmidrule(lr){2-8}
		&\textbf{MASFly} & 98.9  &\textbf{92.9} & \textbf{68.3}& 	\textbf{68.3} &\textbf{65.6} &\textbf{ 61.7}\\
		
		\bottomrule
	\end{tabular}
	\label{tab:tp_performance}
	\vspace{-4px}
\end{table*}

\subsection{Reflective Experience Distillation}
\label{ssec:reflect}

Upon task completion, 
MASFly conducts a reflection process, 
distilling the execution trajectory and evaluation outcomes (on the training set only) into reusable experiences to update the SOP repository and PEP.
For successful executions, the LLM distills the generated OP into a generalized and reusable SOP, 
which is stored in the SOP repository.
For failed executions, MASFly prompts the LLM to  diagnose shortcomings in the OP and attribute errors to specific agents.
These reflective insights serve two purposes: iteratively refining the instantiated OP for successful task completion,
and enriching the PEP to form agent-level priors that guide the process supervision  during future executions.
In this way, 
MASFly progressively strengthens both its system-generation capability and its robustness during execution, leading to increasingly reliable multi-agent coordination over time.
More details are  in Appendix~\ref{appendix:repo}.

\subsection{Comparison with Existing Works}

We compare MASFly with representative methods that do not require extra LLM training in Table~\ref{tab:comparison}.
As shown, MASFly uniquely enables adaptive generation of entire query-dependent systems,
where both agent prompts and communication topology are dynamically customized through the  SOP instantiation mechanism.
It also supports adaptive execution by incorporating an execution-time supervision mechanism.
Moreover, MASFly features an inherent memory mechanism that enables cross-task experience augmentation by reusing successful system-level experience from the SOP repository and agent-level experience of failure from the PEP.
More discussions are in Appendix~\ref{appendix:relatedwork}.

\section{Experiment}

\begin{table*}[ht]
	\centering
	\vspace{-1px}
	\caption{Performance comparison across 4 benchmarks. For methods evaluated over three independent runs, we report the mean; the corresponding standard deviations are provided in Appendix~\ref{appendix:std}.}
	\vspace{-4px}
	\small
	\begin{tabular}{lccccccc}
		\toprule
		\multirow{2}{*}{\textbf{Method}} & \multicolumn{4}{c}{\textbf{GAIA}} & \multirow{2}{*}{\textbf{HumanEval}} & \multirow{2}{*}{\textbf{MBPP}}& \multirow{2}{*}{\textbf{MBPP Pro}}\\
		\cmidrule(lr){2-5} 
		& \textbf{Level 1} & \textbf{Level 2} & \textbf{Level 3} & \textbf{Avg.} &  & &\\
		\midrule
		Base & 29.3 & 11.3 & 0.0 & 16.4 &  94.94 &   91.49 & 82.80 \\
		ReAct & 32.5 & 28.3 & 6.3 & 26.7 & 95.6 & 91.5 & 83.0 \\
		MetaGPT & 34.1 & 31.5 & 6.3 & 28.8 & 96.2 & 90.7 & 82.2\\
		AgentVerse & 30.9 & 20.2 & 8.4 & 22.5 & 96.8 & 90.9 & 81.7 \\
		MegaAgent & 32.9 & 26.4 & 3.1 & 25.5 & 97.15& 92.38 & 83.11\\
		GPTSwarm & 36.6 & 32.1 & 6.3 & 30.0 &  97.78  & 91.79  &84.20\\
		AFlow & 36.6 & 35.8 & 6.3 & 31.8 & 97.78 & 92.96 & 84.74\\
		AgentSquare & 41.5 & 37.7 & 10.4 & 35.1 & 96.8 & 92.1 & 83.9\\
		G-Memory & 43.1 & 37.7 & 12.5 & 36.0 & 96.4 & 92.1 & 83.7 \\\midrule
		\textbf{MASFly} & \textbf{46.3} & \textbf{41.5} & \textbf{16.7} & \textbf{39.7} & \textbf{98.7} & \textbf{93.3} & \textbf{85.6} \\
		\bottomrule
	\end{tabular}
	\label{tab:performance}
	\vspace{-8px}
\end{table*}

\subsection{Experimental Setup}

\paragraph{Benchmarks.}
We evaluate MASFly on five benchmarks involving tool calls covering three domains:
(1)long-horizon planning: \textit{TravelPlanner}~\cite{xie2024travelplanner},which provides two evaluation modes: \textit{Two-Stage} and \textit{Sole-Planning};
(2)general question answering: \textit{GAIA}~\cite{mialon2023gaia}, consisting of questions categorized into three difficulty levels;
and 
(3) code generation: \textit{HumanEval}~\cite{chen2021evaluating}, \textit{MBPP}~\cite{austin2021program}
and \textit{MBPP Pro}~\cite{yu2024humaneval}.
The dataset statistics are in Appendix~\ref{appendix:setup}.
For fair comparison,
MASFly builds repositories offline from a subset of training split and is evaluated on the held-out test split with repositories fixed. 
Baselines that require offline preparation are given the same  data budget.

\paragraph{Baselines.}
We compare MASFly with three series of baselines without model training: 
(1) vanilla single LLMs, including
CoT~\cite{wei2022chain}, ReAct~\cite{yao2022react}, Reflexion~\cite{shinn2023reflexion};
(2) hand-craft MAS, including 
MetaGPT~\cite{hong2023metagpt}; 
(3) (partially or fully) autonomous MASs,
including AgentVerse~\cite{chen2023agentverse}, EvoAgent~\cite{yuan2024evoagent}, GPTSwarm~\cite{zhuge2024gptswarm},
AFlow~\cite{zhang2024aflow}, AgentSquare~\cite{shang2024agentsquare}, MegaAgent~\cite{wang2025megaagent}, SwarmAgentic~\cite{zhang2025swarmagentic}
and G-Memory~\cite{zhang2025g} (with the fixed AutoGen~\cite{wu2024autogen} backbone).
For a fair comparison, all methods use the same Qwen3-235B-A22B-Thinking-2507 backbone~\cite{yang2025qwen3} and are evaluated with the same tool interfaces and sandbox environment.

\paragraph{Implementation details.}
We utilize Qwen3-235B-A22B-Thinking-2507~\cite{yang2025qwen3}
as LLM backbone via API call, with the temperature set to 0.6.
We typically set the number of retrieved SOP $K=2$, and the hyperparameter $\lambda=0.3$.

\subsection{Performance Comparison}
As shown in Table~\ref{tab:tp_performance} and Table~\ref{tab:performance}, MASFly consistently outperforms handcrafted, automated, and memory-augmented multi-agent baselines across all benchmarks. The advantage is particularly pronounced on TravelPlanner, where MASFly improves the final success rate by 62\% and 98\% in two modes, respectively, while maintaining high delivery rates.
This marked improvement validates MASFly’s design, which dynamically instantiates high-quality SOPs that decompose complex tasks and assign tools to specialized agents, and performs effective process supervision and reconfiguration, thereby enabling more coherent and constraint-compliant reasoning in long-horizon planning tasks.
On broader reasoning and programming benchmarks, MASFly also achieves superior performance. 
These consistent improvements demonstrate the contribution of MASFly's in-context adaptability.

\begin{figure}[t]
	\centering
	\vspace{-1px}
	\includegraphics[width=0.46\textwidth]{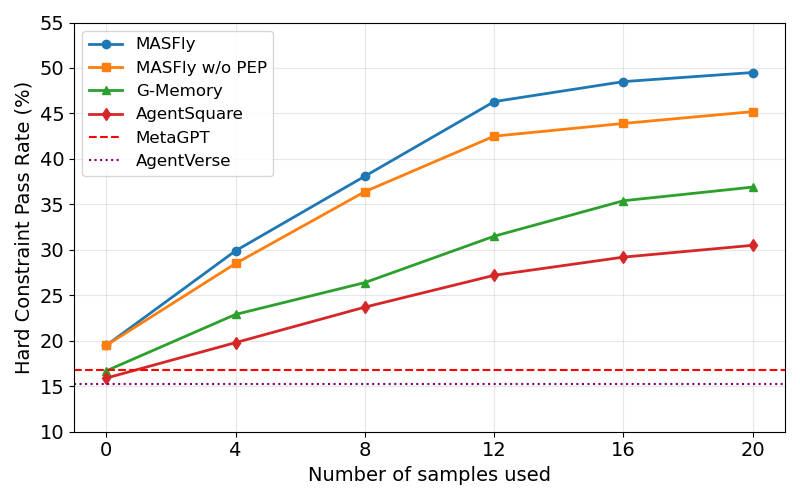}
	\vspace{-4px}
	\caption{Experience Accumulation Curve. }
	\label{fig:lcurve}
	\vspace{-4px}
\end{figure}

We further analyze how MASFly benefits from progressively accumulated experience. Figure~\ref{fig:lcurve} reports the performance on TravelPlanner as more training samples are used by different methods. 
Even in the zero-shot (0 sample in the repository) setting, MASFly outperforms all compared baselines, showing that it can already leverage the LLM’s internal knowledge to support test-time system adaptation, rather than relying solely on fixed static configurations. 
As more samples are incorporated, MASFly improves rapidly and consistently remains above both the search-based baseline AgentSquare and the memory-augmented baseline G-Memory, indicating that the framework can effectively transform accumulated experience into stronger test-time system adaptation, thus outperforming fixed MAS baselines.
We also include a variant without PEP, which shows that distilled failure experience provides additional benefits. Overall, the curve demonstrates both the effectiveness of progressive experience accumulation and the strong sample efficiency of MASFly. 
More efficiency comparisons of different methods are provided in Appendix~\ref{appendix:effi}.

\begin{figure}[t]
	\centering
	\vspace{-1px}
	\includegraphics[width=0.46\textwidth]{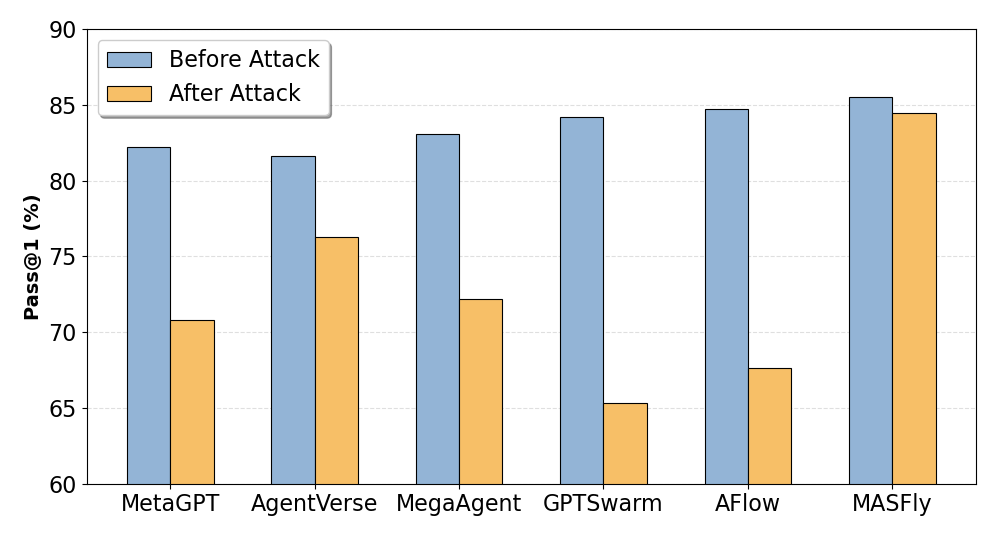}
	\vspace{-4px}
	\caption{The performances before and after prompt attack on MBPP Pro. }
	\label{fig:attack}
	\vspace{-4px}
\end{figure}

\begin{figure*}[ht]
	\centering
	\includegraphics[width=0.94\textwidth]{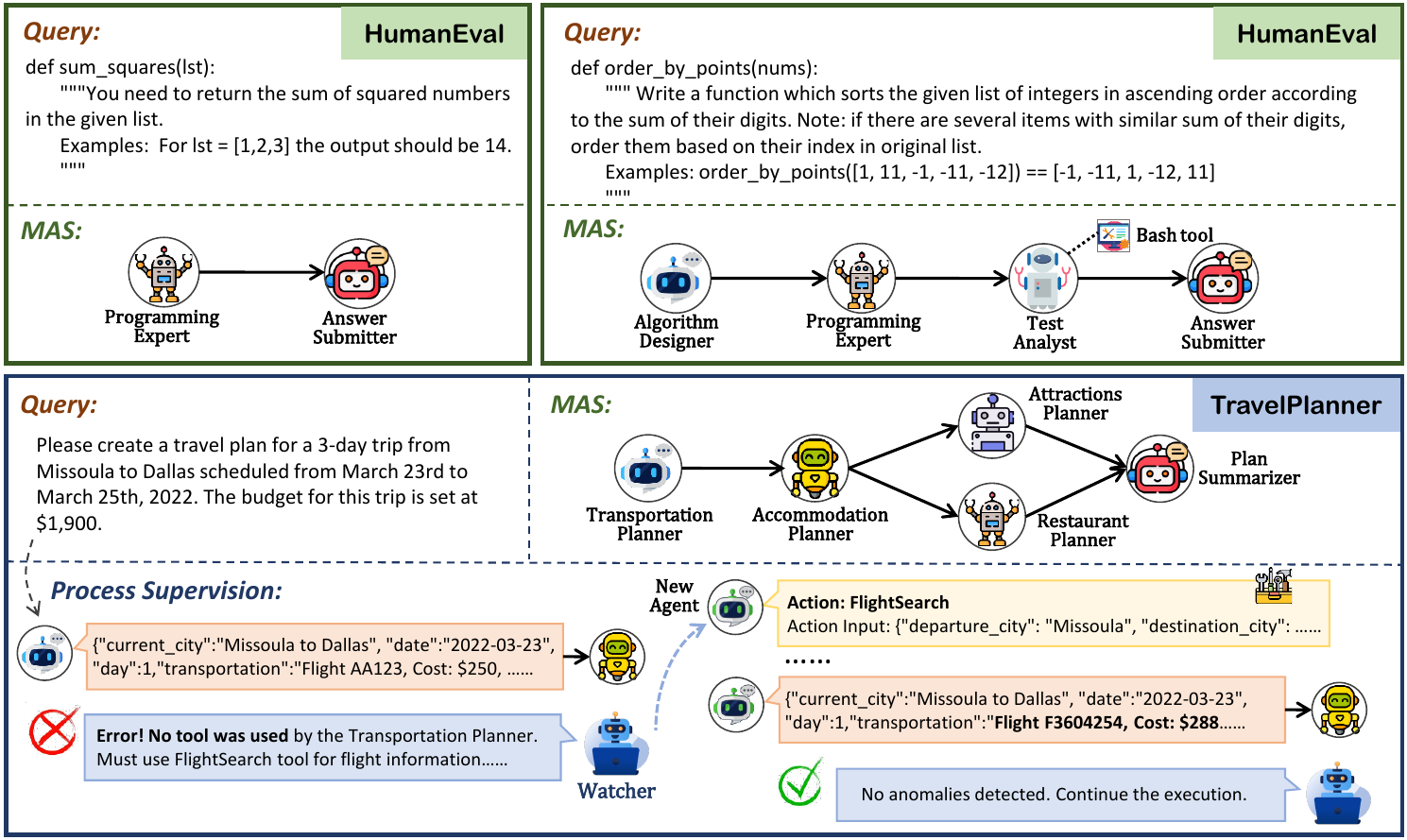}
	\vspace{-8px}
	\caption{Case studies of adaptive system generation and process supervision in MASFly. }
	\label{fig:case}
	\vspace{-4px}
\end{figure*}

\subsection{Robustness Analysis}

To assess the robustness of MAS under adversarial conditions, we follow prior work~\cite{zhuge2024gptswarm} and introduce a prompt attack during test time on MBPP Pro benchmark, where we force the coding agent to output an empty function.
Figure~\ref{fig:attack} reports the performance before and after attack across representative systems. 
Most methods experience a substantial degradation after attack, indicating their inability to mitigate abnormal agent behaviors once execution begins.
In contrast, MASFly maintains almost the same performance. 
This robustness stems from MASFly’s process supervision: the Watcher identifies the compromised agent, removes it and dynamically instantiates a replacement agent to continue  execution.
These results highlight how MASFly’s execution-time adaptation helps preserve high task accuracy under adversarial or abnormal conditions.

\subsection{Ablation Study}
\label{ssec:ablation}

\paragraph{Effect of Adaptive System Generation.}
As shown in Table~\ref{tab:ablation}, adaptive system generation is crucial to the performance. 
Replacing SOP-guided instantiation with a handcrafted MAS configuration (\textit{w/o SOP-RAG}) leads to the largest drop, indicating that fixed systems cannot match the diverse demands of different queries. 
Using the retrieved SOP directly as the system configuration (\textit{w/o SOP-Inst}) also degrades performance, showing that retrieval alone is insufficient without further query-specific instantiation. 
These results confirm that MASFly benefits not only from retrieving relevant collaboration priors, but also from adapting them into a tailored system for the current query.

\begin{table}[t]
	\centering
	\caption{Comparison of different variants of MASFly.}
	\vspace{-4px}
	\setlength\tabcolsep{6pt}
	\small
	\begin{tabular}{lcccc}
		\toprule
		\multirow{2}{*}{Method} & \multicolumn{2}{c}{TravelPlanner} & \multicolumn{2}{c}{GAIA}\\
		\cmidrule(lr){2-3} \cmidrule(lr){4-5} 
		& Common & Hard & Level 1 & Level 2  \\
		\midrule
		MASFly & \textbf{85.3} &\textbf{49.5} &\textbf{46.3} & \textbf{41.5}  \\
		\midrule
		w/o SOP-RAG & 65.3 & 18.6 &  35.4  &  32.1 \\
		w/o SOP-Inst & 80.8  &  40.9  &  41.4  & 35.8 \\
		w/o Watcher  & 83.9  & 37.1  &  41.4 & 36.8 \\
		w/o PEP &  84.4 &  45.2 & 43.3 & 39.6\\
		\bottomrule
	\end{tabular}
	\label{tab:ablation}
	\vspace{-4px}
\end{table}

\paragraph{Effect of Adaptive Execution.}

Table~\ref{tab:ablation} also shows the importance of adaptive execution. 
Removing the Watcher agent (\textit{w/o Watcher}) causes a clear drop, as the system loses execution-time reconfiguration ability to correct abnormal behaviors. 
Moreover, disabling the personalized experience pool (\textit{w/o PEP}) still degrades performance, indicating that accumulated experience provides valuable guidance for more effective interventions. 
These results show that execution-time supervision and experience-guided intervention work together to improve robustness and stability.

\subsection{Case Study}

To further illustrate how MASFly dynamically adapts to query requirements and ensures reliable execution, we present several representative cases in Figure~\ref{fig:case}.
The first two examples from HumanEval demonstrate how MASFly autonomously constructs adaptive MAS for different queries.
Simpler problems lead to a compact two-agent system, while more complex ones trigger a four-agent configuration with tools.
This adaptive system generation is guided by the retrieved SOP templates, while the LLM’s internal reasoning ensures appropriate role assignment.
The third example from TravelPlanner shows that 
during execution, the Watcher detects that the first planner fabricates flight information without invoking the tool.
Then Watcher intervenes, reconfigures it with a new agent, and urges it to follow the correct procedure and retrieve flight data via the tool.
This demonstrates how process supervision enables real-time error detection and behavioral correction, ensuring the reliability and adaptability of the overall system.

\section{Conclusion}
In this paper, we formulate in-context structural adaptation for LLM-based MAS and introduce MASFly as a concrete realization.
MASFly performs query-dependent system generation through retrieval-augmented SOP instantiation and execution-time reconfiguration through experience-enhanced process supervision.
Experiments show that MASFly delivers strong performance with high adaptability and robustness, suggesting that structured context can adapt not only the behavior of an individual LLM but also the organization through which multiple LLM agents solve a task.
 


\section*{Limitations}


One limitation of MASFly is that the triggering mechanism of the process supervision relies on pre-defined heuristic rules (e.g., setting the intervention interval to half the number of agents) to initiate interventions. 
This approach may not be optimal for all scenarios. 
In future work, it is worth exploring automated anomaly detection mechanisms that allow the system to autonomously determine the precise timing for intervention, thereby reducing the reliance on manual heuristics.
Moreover, constructing the SOP repository and personalized experience pool requires labeled training instances, and experience collected from limited domains may introduce bias or negative transfer when reused in substantially different settings.
The Watcher also has incomplete visibility into some agent--environment interactions and may issue unnecessary interventions, while supervision and reconfiguration introduce additional inference cost.
Future work should therefore investigate principled experience selection, more reliable intervention policies, and budget-aware adaptation.

\section*{Acknowledgments}

This work is supported by Ant Group through CCF-Ant Research Fund
and Beijing Science and Technology Program (No.Z251100008125003).

\bibliography{custom}

\clearpage
\appendix

\section{Comparison with Related Works}
\label{appendix:relatedwork}

In Table~\ref{tab:comparison}, we compare MASFly with representative MAS from multiple perspectives.
Here, in-context adaptation covers query-dependent system generation and execution-time reconfiguration without updating LLM parameters.
Adaptive system generation entails tailoring agent prompts and communication topologies to each input query, while adaptive execution involves dynamically adjusting system configurations throughout the execution phase.
Cross-task memory refers to the system's capability to reuse memories or experiences from previous tasks or queries to facilitate the current task.
Specifically:
\begin{itemize}[leftmargin=*]
	\item MetaGPT~\cite{hong2023metagpt} is a pioneering framework for constructing LLM-based MAS with diverse roles. However, it relies on fixed, handcrafted agent prompts and communication structures, limiting adaptability to different queries. In addition, it lacks process supervision and experience reuse mechanisms.
	Similar works include ChatDev~\cite{qian2023chatdev} and AutoGen~\cite{wu2024autogen}.
	\item AgentVerse\cite{chen2023agentverse} leverages an LLM to automatically generate diverse agent roles based on user queries, enhancing prompt adaptability. However, it still employs a static communication structure, making it not a fully adaptive system. It can dynamically replace team members by self-assessing execution results, but it lacks experience augmentation mechanism. Similar works include EvoAgent\cite{yuan2024evoagent}.
	\item MegaAgent~\cite{wang2025megaagent} prompts a Boss agent to customize agent roles and does not predefine communication structures, relying instead on agents to coordinate themselves during execution. However, this approach may lead to inefficiency and confusion in long-horizon tasks without explicit communication design. It introduces a hierarchical monitoring mechanism, where internal validation occur between agents, rather than having an external, global Watcher like MASFly to provide unified process supervision, including agent communication and environment interaction monitoring. Moreover, it lacks mechanisms for experience reuse.
	\item GPTSwarm~\cite{zhuge2024gptswarm} focuses on optimizing communication structures among agents through adaptive communication graphs. However, it does not adapt agent prompts for different queries. It also cannot adapt to or handle unexpected situations during testing, and lacks a mechanism for reusing past experience. Similar works include G-Designer~\cite{zhang2024g}.
	\item AgentSquare~\cite{shang2024agentsquare} does not involve multi-role playing. Instead, it uses an LLM to search for an optimal agentic system on a training dataset, including modules for planning, tool usage and memory. However, it adopts a fixed linear sequential collaboration structure. Additionally, it lacks dynamic adaptability after deployment, and lacks experience reuse mechanisms. Similar works include ADAS~\cite{hu2024automated}, AFlow~\cite{zhang2024aflow}, MaAS~\cite{zhang2025multi}, and SwarmAgentic~\cite{zhang2025swarmagentic}.
\end{itemize}

Beyond these methods, there are also recent training-based approaches, such as MAS-GPT\cite{ye2025mas} and FlowReasoner\cite{gao2025flowreasoner}, which train a meta-agent using supervised fine-tuning (SFT) or reinforcement learning (RL) to adaptively generate MAS based on different queries. However, these methods require large amounts of training data and computational resources, and they also lack execution adaptability after deployment and experience reuse mechanisms.
Additionally, memory-based methods (e.g., G-Memory~\cite{zhang2025g})) design specific memory modules to retrieve and reuse interaction logs and experiences from prior similar tasks. Yet, relying on fixed MAS architectures renders them incapable of adaptive system generation and dynamic reconfiguration.

In contrast, MASFly does not update LLM parameters and enables adaptive generation of entire query-dependent systems.
This is achieved through the retrieval-augmented SOP instantiation mechanism, where both agent prompts and communication topology are dynamically customized.
It also supports adaptive execution by incorporating a process supervision and optimization mechanism.
Moreover, MASFly features an inherent memory mechanism that enables experience augmentation by reusing successful system-level experience from the SOP repository and agent-level failure experience from the PEP.


\section{Implementation Details}

\subsection{Details of the Repository}
\label{appendix:repo}

\subsubsection{Repository Initialization}

To initialize the SOP repository, we randomly sample a subset of training instances and construct an initial collection of SOP cases.
For the TravelPlanner and GAIA benchmarks, where multiple tools are available, we encourage the LLM (Qwen3-235B) to generate diverse agent roles, each responsible for distinct tools, together with a dedicated final-answer submitter to ensure response quality.
For coding benchmarks such as HumanEval, queries are often simple and the Bash tool is unnecessary. To maintain system efficiency, we prompt the LLM to begin by designing a minimal MAS (e.g., a single-agent team). If the resulting solution is incorrect, the LLM is then guided to incrementally introduce additional roles to enhance execution quality.
This staged construction strategy yields an initial repository that is both diverse and computationally efficient.

\subsubsection{Repository Update}

Updates to both the SOP repository and the Personalized Experience Pool (PEP) are carried out during the third stage of our method, reflective experience distillation.
Note that this process takes place \textbf{only on labeled training data}, not on the test set.
During testing, ground truth is not visible to the system; thus, no iterative reflective correction is performed.

For successful executions, the LLM is prompted to summarize the current task-specific OP into a more abstract SOP.
This involves converting agent instructions that rely heavily on the user’s query into generalized, reusable directives.
During this process, the LLM also references previously retrieved SOPs, extracting and reusing valuable instructions.
The resulting SOP template is then stored in the SOP repository.

For failed executions, the LLM receives the detailed evaluation outcome together with the full MAS execution trajectory (i.e., communication logs and tool-use traces).
It then performs reflection to diagnose the failure, attribute individual agent errors, and update the personalized experience pool accordingly.
The reflective feedback is also used to revise the task’s OP—such as adjusting agent roles, modifying the communication structure, or refining agent instructions and tool usage—and the task may be re-executed until success or until the maximum number of iterations is reached.

\paragraph{Qualitative evolution of the repository.}
We further inspect representative SOPs collected at different stages of repository construction.
As more training experience is incorporated, the repository evolves from generic collaboration patterns toward more specialized role decomposition, tool allocation, communication structures, and agent instructions.
Personalized experience distilled from failures helps refine agent instructions by identifying recurring errors and their corresponding improvement strategies, while successfully revised workflows are subsequently distilled into reusable SOPs.
For example, later-stage SOPs may introduce more explicit testing roles for coding tasks or more detailed tool-use and constraint-checking instructions for TravelPlanner.
This qualitative trend is consistent with the experience accumulation results in Figure~\ref{fig:lcurve}, where performance improves as the repository becomes richer and its experience becomes more reusable.

Because SOPs are compact, high-level, and query-agnostic, only a small amount of training data is needed to build the SOP repository and the corresponding PEP.
In practice, the final SOP repository contains fewer than 100 entries across all tasks (see Table~\ref{tab:dataset}).
Therefore, no additional deletion or curation strategy is required.
If future applications involve significantly more diverse or complex scenarios, lightweight scalability can be achieved by clustering SOP embeddings and retaining cluster centroids to maintain a compact repository.

\subsection{Agent Communication Details}

To transcend the constraints of static blueprints and enhance adaptability, MASFly adopts a {bidirectional communication mechanism}: each agent can send messages to its downstream agents as prescribed by $OP$, but can also communicate upward to the agents from which it previously received messages—for example, requesting clarification or more complete information. 

The agent’s output format adheres to the ReAct~\cite{yao2022react} structure, meaning it first outputs a thought, followed by an action, which may involve either using a tool or sending a message to another agent.
All inter-agent messages are stored in a global message pool, which enables efficient and parallel processing of communication events, ensuring both scalability and consistency during execution.
All tool invocations by agents are logged into tool usage history, allowing Watcher to review. 
When the Watcher decides to remove an anomalous agent, all related entries of that agent in the message pool and the tool usage history are completely purged, thereby thoroughly eliminating the impact of the anomalous agent.

\subsection{Process Supervision Details}

To guide process supervision, we retrieve relevant past failure experiences from the PEP for the Watcher's reference. For the sake of efficiency, this is conducted as a single retrieval step before the task begins. We utilize the hybrid scores computed during SOP retrieval to fetch PEP cases associated with the same query index. 
Given the concise nature of the agent-wise experiences (as shown in Figure~\ref{fig:Personrepo}), we typically retrieve about two similar cases to form the Watcher's reference context. Empirically, this retrieved content accounts for approximately 30\% of the Watcher's total input prompt. Leveraging the strong context capabilities of advanced LLMs, we directly append these cases into the prompt, which provides sufficient guidance without causing context pollution. 
If a more lightweight model is deployed as the Watcher, flexible context-management strategies could be adopted—such as introducing a preliminary filtering step to extract the most pertinent details before passing them to the Watcher. 
Furthermore, the ablation study in Section~\ref{ssec:ablation} effectively demonstrates the performance benefits of incorporating the PEP..

During runtime, the Watcher is triggered periodically (i.e., every \(M\) communication rounds, every \(L\) environment-interaction steps) to monitor the system. 
When triggered, it reviews the agents' environmental interaction logs and inter-agent communication histories, comparing their behaviors against the prescribed OP and the retrieved PEP priors. 
On the one hand, the Watcher performs routine checks, such as ensuring agents do not fall into endless loops, engage in unproductive communication, or violate the designated OP. On the other hand, it conducts task-specific inspections by drawing upon the PEP cases to identify potential task-level issues, such as tool misuse or constraint violations.
When the Watcher observes that an agent is caught in unproductive environmental interactions, it clears the interaction history and sends experience-guided feedback to correct the agent's strategy. 
Alternatively, if abnormal inter-agent communication is detected, the Watcher intervenes by removing the anomalous agent and its logs, and dynamically re-instantiates a new agent to resume the workflow. 
This targeted supervision guarantees the robust execution of the system.

\subsection{Experimental Setup}
\label{appendix:setup}

\subsubsection{Datasets and Evaluation Metrics}
Following~\citet{zhuge2024gptswarm,wang2025megaagent}, we evaluate MASFly and compared methods on five representative benchmarks involving tool calls across three domains:
\begin{itemize}[leftmargin=*]
	\item \textbf{Long-horizon planning:} \textit{TravelPlanner}~\cite{xie2024travelplanner} includes two modes: \textit{two-stage} and \textit{sole-planning}. We mainly focus on the two-stage mode, where the model is required to use six tools to gather information from a sandbox environment and formulate a complete travel plan.
	Since most previous methods have not adopted this more challenging setting but rather the sole-planning mode, where all necessary information is provided without tool use, we also evaluate our method under this easier mode for completeness.
	Following the original evaluation setting, the evaluation metrics include delivery rate, commonsense constraint pass rate, hard constraint pass rate, and final plan pass rate.
	
	\item \textbf{General QA:} \textit{GAIA}~\cite{mialon2023gaia} is a benchmark specifically designed to test the generality of AI assistants on real-world, knowledge-intensive questions.  
	Answering GAIA questions typically requires a combination of reasoning, web searching, and tool utilization.  
	Following~\cite{zhuge2024gptswarm}, all methods are provided only with the Google Search tool via SerpAPI\footnote{\url{https://serpapi.com/}} and a local file processing tool, without access to any paid browsing APIs or multimodal tools.  
	Accuracy is used as the evaluation metric.  
	\item \textbf{Coding:}  
	(1) \textit{HumanEval}~\cite{chen2021evaluating} is a hand-crafted benchmark designed to assess code generation capabilities in Python.  
	Each problem includes a function signature and several examples that specify the intended functionality.  
	
	(2) \textit{MBPP}~\cite{austin2021program} is a large-scale dataset containing short, human-written Python programming problems covering diverse skills.  
	Each problem includes a function signature and one example test case sampled from the dataset.  
	
	(3) \textit{MBPP Pro}~\cite{yu2024humaneval} is an expanded versions of the MBPP benchmarks.
	Considering the issue of incorrect answer labeling in the MBPP dataset that hinder performance improvement, we consider the more challenging MBPP Pro dataset to validate the method's performance.
	
	For fair comparison, all test cases are held out during system execution and are only used for post-hoc correctness verification.  
	For both datasets, the Bash tool is enabled for running LLM-generated test cases and analyzing execution results.  
	Pass@1 is adopted as the evaluation metric. 
	We report the mean results from three independent runs. 
\end{itemize}

The dataset statistics are summarized in Table~\ref{tab:dataset}, where we sample instances from the training sets to construct the SOP repository.

\begin{table}[t]
	\centering
	\caption{Statistics of the evaluated datasets.}
	\vspace{-8px}
	\begin{tabular}{lcc}
		\toprule
		\textbf{Dataset} & \textbf{\#Train} & \textbf{\#Test} \\
		\midrule
	{TravelPlanner} & 20 & 180 \\
	{GAIA (Level 1)} & 5 & 41 \\
	{GAIA (Level 2)} & 3 & 53 \\
	{GAIA (Level 3)} & 2 & 16 \\
	{HumanEval} & 6 & 158 \\
	{MBPP} & 20 & 341 \\
	{MBPP Pro} & 10 & 367 \\
		\bottomrule
	\end{tabular}
	\label{tab:dataset}
\end{table}

\subsubsection{Baselines}

We compare MASFly with three series of baselines without additional training: 
\begin{itemize}[leftmargin=*]
	\item CoT~\cite{wei2022chain}: A prompting strategy that enhances problem-solving by guiding the LLM to generate a series of intermediate reasoning steps before arriving at the final answer.
	\item ReAct~\cite{yao2022react}: A framework that interleaves reasoning traces with action execution, enabling the LLM to interact with  environments to retrieve information.
	\item Reflexion:~\cite{shinn2023reflexion}: A framework that enhances the reasoning process by incorporating a self-reflective mechanism, enabling agents to iteratively evaluate and refine their own thought trajectories.
	\item MetaGPT~\cite{hong2023metagpt}: A multi-agent framework that mimics a software company by assigning agents to specific roles and regulating their collaboration through hand-crafted SOPs.  
	For non-programming tasks, we select a fixed SOP generated by the LLM to construct the system.
	\item AgentVerse~\cite{chen2023agentverse}: A multi-agent framework that uses an LLM to dynamically generate diverse agent roles based on the user query and facilitates collaboration through a multi-turn discussion and self-evaluation process.
	\item EvoAgent~\cite{yuan2024evoagent}: A method that leverages evolutionary algorithms to automatically generate diverse agent roles.
	\item MegaAgent~\cite{wang2025megaagent}: A hierarchical framework where a central "Boss" agent autonomously customizes sub-agent roles and coordinates them to handle complex, long-horizon tasks.
	\item GPTSwarm~\cite{zhuge2024gptswarm}: An approach that optimizes the communication topology between agents by treating the system structure as a learnable graph to improve information flow.
	\item AFlow~\cite{zhang2024aflow}: An automated framework that formulates the construction of agentic workflows as a code search problem, using algorithms of MCTS to find the optimal structure.
	\item AgentSquare~\cite{shang2024agentsquare}: A search-based method that identifies an optimal system configuration by searching within a predefined design space containing modules for planning, tool usage, and memory.
	\item SwarmAgentic~\cite{zhang2025swarmagentic}: A framework inspired by Particle Swarm Optimization that automatically constructs and evolves a population of agentic systems through search. We reproduce it using the recently released official code under our unified evaluation setting.
	\item G-Memory~\cite{zhang2025g} is a hierarchical memory system for MAS, which retrieves and reuses the historically relevant MAS interaction logs and experience. Following the setting of original paper, we integrate it with AutoGen~\cite{wu2024autogen} framework.
\end{itemize}

For a fair comparison, all methods use the same Qwen3-235B-A22B-Thinking-2507 backbone~\cite{yang2025qwen3} and are evaluated with the same tool interfaces and sandbox environment.

\subsubsection{Hyperparameters}

We utilize Qwen3-235B-A22B-Thinking-2507~\cite{yang2025qwen3} as the LLM backbone via API calls, with the temperature set to 0.6. For the system generation phase, we typically set the number of retrieved SOP candidates 
$K=2$ and the weight hyperparameter $\lambda=0.3$. 
Regarding the Watcher intervention mechanism during execution, we adopt default heuristic settings to initiate supervision: the intervention interval for communication rounds $M$ is set to half the number of agents, and the threshold for consecutive environment interaction steps 
$L$ is set to 5.

\section{Complementary Experiments}

\subsection{Generalization across Different Backbone}

\begin{table*}[ht]
	\centering
	\caption{Performance comparison on TravelPlanner Two-Stage mode using Qwen3-32B.}
	\vspace{-8px}
	\small
	\begin{tabular}{lcccccc}
		\toprule
		\multirow{2}{*}{\textbf{Method}} & \multirow{2}{*}{\textbf{Delivery Rate}} & \multicolumn{2}{c}{\textbf{Commonsense}} & \multicolumn{2}{c}{\textbf{Hard Constraint}} & \multirow{2}{*}{\textbf{Final Pass Rate}} \\
		\cmidrule(lr){3-4} \cmidrule(lr){5-6}
		&  & \textbf{Micro} & \textbf{Macro} &\textbf{Micro} &\textbf{Macro} &  \\
		\midrule
		ReAct & 70.5 & 48.8 & 2.8 & 5.7 & 2.8 & 0 \\
		MetaGPT & \textbf{100.0} & 56.1 & 7.2 & 12.1 & 4.4 & 5.6 \\
		AgentVerse & \textbf{100.0} & 52.7 & 4.8 & 6.2 & 4.4 & 2.2 \\
		AgentSquare &\textbf{100.0} & 61.4 & 11.7 & 15.2 & 11.1& 8.3 \\ 
		G-Memory &  \textbf{100.0} & 66.0 & 14.4 & 15.2 & 13.9 & 9.4 \\ \midrule
		\textbf{MASFly} & \textbf{100.0} & \textbf{68.3} & \textbf{15.6} & \textbf{19.0} & \textbf{17.8} & \textbf{13.3} \\
		\bottomrule
	\end{tabular}
	\label{tab:32B_performance}
\end{table*}

To evaluate whether MASFly generalizes beyond large-capacity LLMs, we replace the backbone with a lighter Qwen3-32B model~\cite{yang2025qwen3} and test under the challenging Two-Stage setting of TravelPlanner. 
As shown in Table~\ref{tab:32B_performance}, MASFly still achieves clear performance gains over both manually designed and automated MAS baselines. 
This demonstrates that our framework does not rely on a specific or extremely large backbone; instead, the structured SOP-guided design and supervision mechanism enable strong robustness and generalization across different model scales.

\subsection{Results across Independent Runs}
\label{appendix:std}

\begin{table*}[t]
	\centering
	\caption{Results over three independent runs, reported as mean $\pm$ standard deviation (\%).}
	\label{tab:std_results}
	\vspace{-4px}
	\small
	\begin{tabular}{lcccccc}
		\toprule
		\textbf{Method} & \textbf{GAIA-L1} & \textbf{GAIA-L2} & \textbf{GAIA-L3} & \textbf{HumanEval} & \textbf{MBPP} & \textbf{MBPP Pro} \\
		\midrule
		ReAct       & $32.5 \pm 1.4$ & $28.3 \pm 1.9$ & $6.3 \pm 0.0$  & $95.6 \pm 0.0$ & $91.5 \pm 0.0$ & $83.0 \pm 0.2$ \\
		MetaGPT     & $34.1 \pm 0.0$ & $31.5 \pm 1.1$ & $6.3 \pm 0.0$  & $96.2 \pm 0.0$ & $90.7 \pm 0.2$ & $82.2 \pm 0.2$ \\
		AgentVerse  & $30.9 \pm 1.4$ & $20.2 \pm 1.1$ & $8.4 \pm 3.6$  & $96.8 \pm 0.6$ & $90.9 \pm 0.3$ & $81.7 \pm 0.3$ \\
		AgentSquare & $41.5 \pm 2.5$ & $37.7 \pm 1.9$ & $10.4 \pm 3.6$ & $96.8 \pm 0.0$ & $92.1 \pm 0.3$ & $83.9 \pm 0.3$ \\
		G-Memory    & $43.1 \pm 1.4$ & $37.7 \pm 1.9$ & $12.5 \pm 0.0$ & $96.4 \pm 0.3$ & $92.1 \pm 0.3$ & $83.7 \pm 0.2$ \\
		\midrule
		\textbf{MASFly} & $\mathbf{46.3 \pm 2.5}$ & $\mathbf{41.5 \pm 1.9}$ & $\mathbf{16.7 \pm 3.6}$ & $\mathbf{98.7 \pm 0.6}$ & $\mathbf{93.3 \pm 0.3}$ & $\mathbf{85.6 \pm 0.3}$ \\
		\bottomrule
	\end{tabular}
\end{table*}

For GAIA and the coding tasks, we conduct three independent runs of MASFly and five major baselines.
As shown in Table~\ref{tab:std_results}, MASFly achieves the highest mean performance on every evaluated benchmark.
The larger variance on GAIA Level 3 is expected because its test split contains only 16 examples, where a single prediction changes accuracy by 6.25 percentage points.
Overall, the repeated-run results support the robustness of the main performance trend.

\subsection{Task-Level Bootstrap Confidence Intervals on TravelPlanner}
\label{appendix:tp_bootstrap}

\begin{table}[t]
	\centering
	\caption{Final Pass Rate and task-level 95\% bootstrap confidence intervals on TravelPlanner.}
	\label{tab:tp_bootstrap}
	\vspace{-4px}
	\small
	\setlength\tabcolsep{6pt}
	\begin{tabular}{clcc}
		\toprule
		\textbf{Mode} & \textbf{Method} & \textbf{Pass Rate} & \textbf{95\% CI} \\
		\midrule
		\multirowcell{7}{\textit{Two-}\\\textit{Stage}}
		& ReAct       & 5.6  & [2.2, 8.9] \\
		& Reflecxion  & 7.8  & [3.9, 11.7] \\
		& MetaGPT     & 8.9  & [5.0, 13.3] \\
		& AgentVerse  & 2.2  & [0.6, 4.4] \\
		& AgentSquare & 19.4 & [13.9, 25.6] \\
		& G-Memory    & 24.4 & [18.3, 31.1] \\
		\cmidrule(lr){2-4}
		& \textbf{MASFly} & \textbf{39.4} & \textbf{[32.2, 46.7]} \\
		\midrule
		\multirowcell{7}{\textit{Sole-}\\\textit{Planning}}
		& CoT          & 6.7  & [3.3, 10.6] \\
		& MetaGPT      & 8.3  & [4.4, 12.8] \\
		& EvoAgent     & 8.3  & [4.4, 12.8] \\
		& AgentVerse   & 6.7  & [3.3, 10.6] \\
		& MegaAgent    & 10.0 & [6.1, 14.4] \\
		& SwarmAgentic & 31.1 & [24.4, 37.8] \\
		\cmidrule(lr){2-4}
		& \textbf{MASFly} & \textbf{61.7} & \textbf{[54.4, 68.9]} \\
		\bottomrule
	\end{tabular}
\end{table}

Since TravelPlanner requires costly long-horizon interactions, we conduct one full-set run and quantify task-level sampling uncertainty using non-parametric bootstrap.
For each method, we resample the 180 test instances with replacement and recompute the Final Pass Rate over 200,000 bootstrap samples.
We report the 2.5th and 97.5th percentiles as the 95\% confidence interval.

\subsection{Efficiency Comparison}
\label{appendix:effi}

\begin{figure}[ht]
	\centering
	\begin{subfigure}{0.95\columnwidth}
		\centering
		\includegraphics[width=\linewidth]{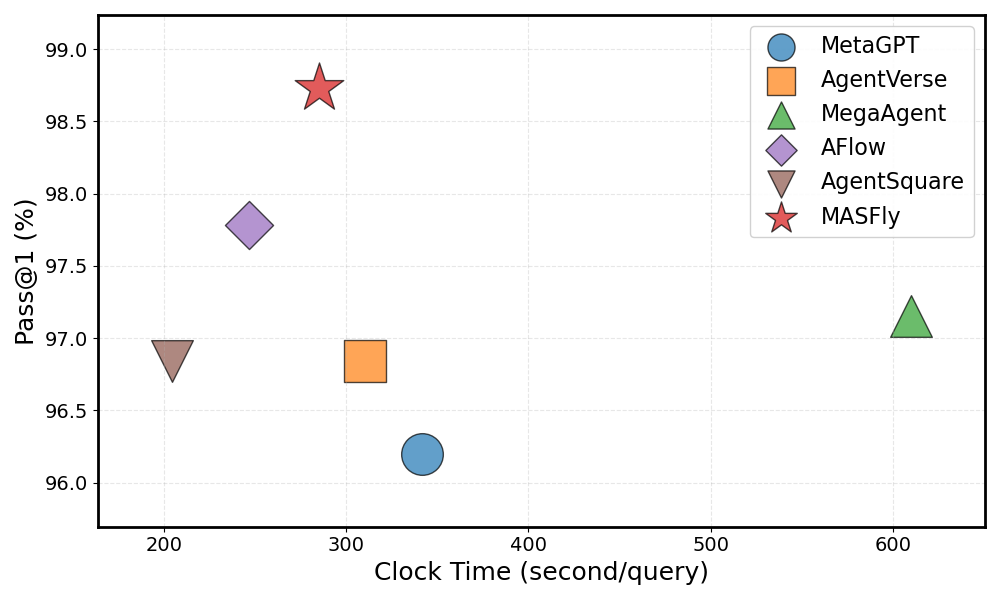}
		\caption{Clock Time vs. Pass@1 on HumanEval.}
	\end{subfigure}
	\vspace{0.5em}
	\begin{subfigure}{0.95\columnwidth}
		\centering
		\includegraphics[width=\linewidth]{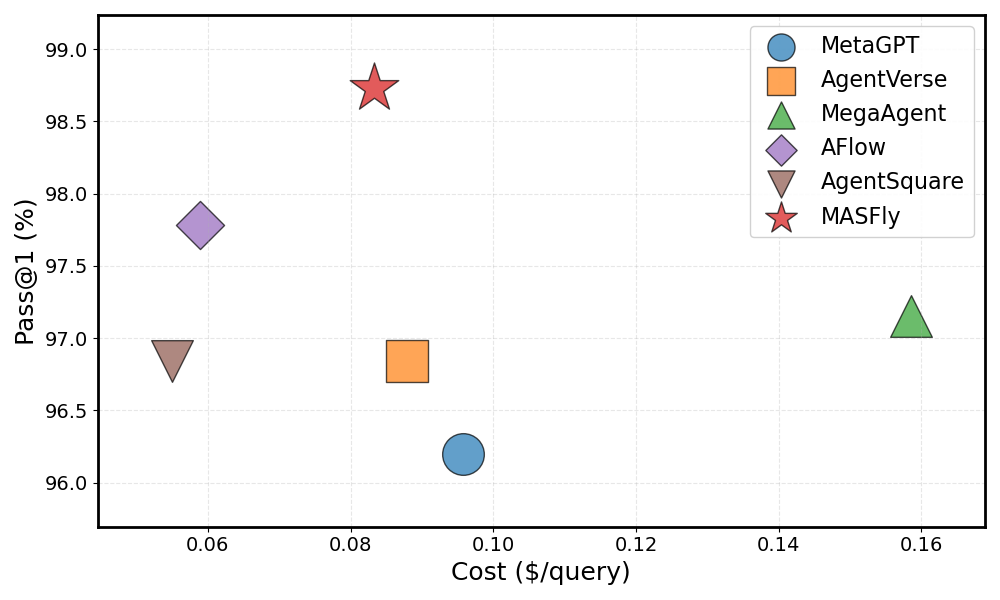}
		\caption{API Cost vs. Pass@1 on HumanEval.}
	\end{subfigure}
	\vspace{-8px}
	\caption{Efficiency–performance comparisons on HumanEval.}
	\label{fig:efficiency}
	\vspace{-10px}
\end{figure}

\begin{figure}[ht]
	\centering
	\begin{subfigure}{0.95\columnwidth}
		\centering
		\includegraphics[width=\linewidth]{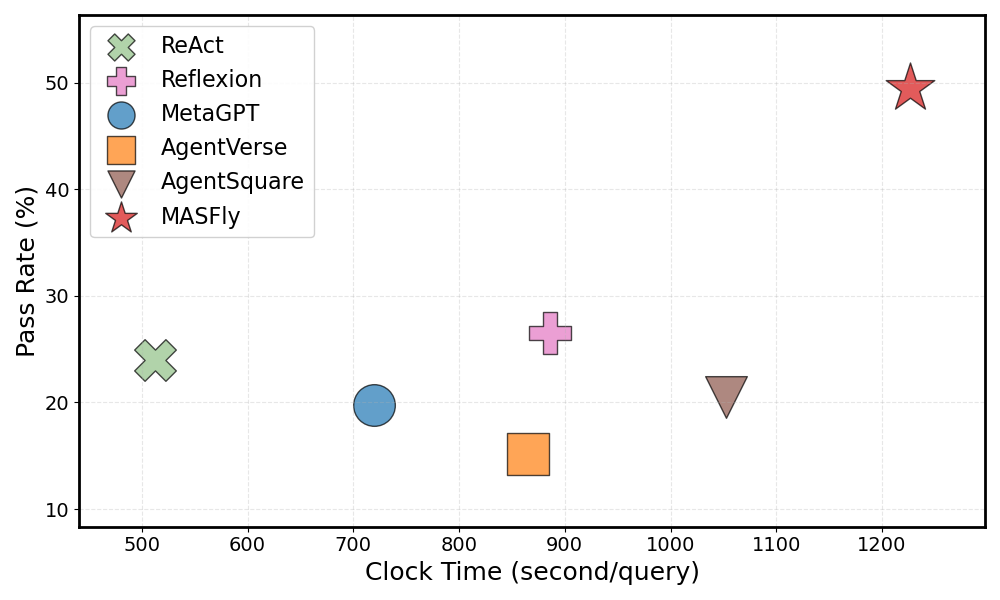}
		\caption{TravelPlanner results.}
		\label{fig:tp}
	\end{subfigure}
	\vspace{0.5em}
	\begin{subfigure}{0.95\columnwidth}
		\centering
		\includegraphics[width=\linewidth]{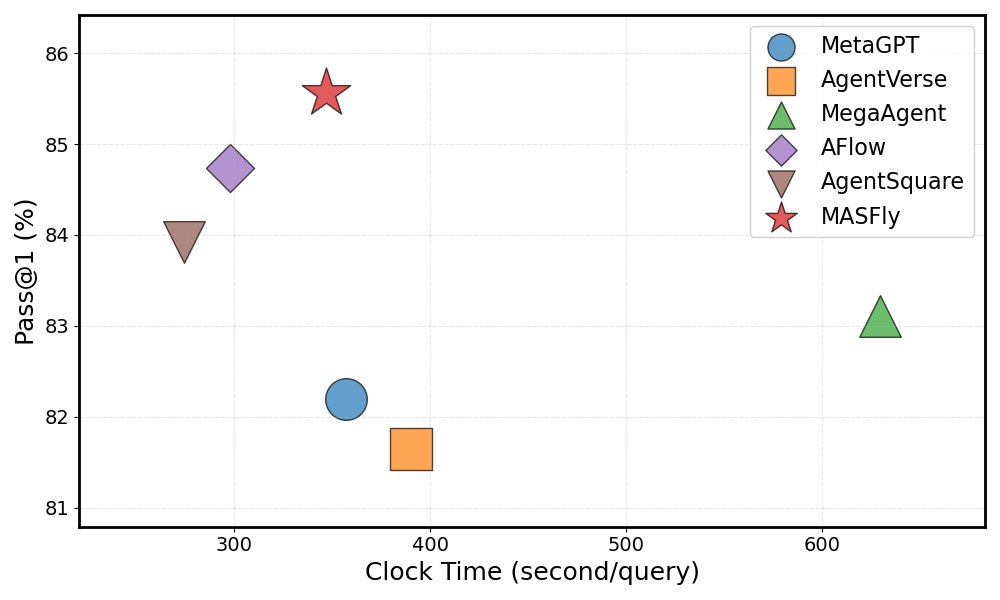}
		\caption{MBPP-Pro results.}
		\label{fig:mbpp}
	\end{subfigure}
	\vspace{-8px}
	\caption{Efficiency–performance comparisons on additional datasets.}
	\label{fig:appendix_efficiency}
	\vspace{-10px}
\end{figure}

We evaluate both efficiency and performance of representative methods on the HumanEval benchmark,
where we measure the clock time (including model inference and API communication latency) and API cost for inference-time efficiency comparison.
Notably, the clock time consists of both system generation time and system execution time.
Figure~\ref{fig:efficiency} shows that MASFly achieves the best overall balance among all methods: it attains the highest Pass@1 accuracy while maintaining moderate clock time and API cost.
This places MASFly on the Pareto frontier of the efficiency–performance trade-off.
The advantage mainly stems from its ability to dynamically customize systems of different complexity based on the given query, improving reasoning quality without incurring excessive computational overhead.

We further report efficiency–performance plots on the TravelPlanner and MBPP-Pro datasets in Figures~\ref{fig:tp} and~\ref{fig:mbpp}.
Across both datasets, MASFly consistently delivers strong performance, while its clock time varies depending on the complexity of the generated multi-agent system.
On TravelPlanner, baseline methods achieve shorter clock time.
However, they often hallucinate intermediate facts and skip interactions with the environment, reducing computation but significantly harming correctness and leading to low hard-constraint pass rates.
In contrast, MASFly spends additional time on supervision and correction stages to validate intermediate outputs and ensure consistency with the database.
This additional verification leads to substantially higher pass rates while keeping the computation budget reasonable.
On MBPP-Pro, tasks are generally more challenging than in HumanEval.
To handle the increased reasoning demand, MASFly sometimes generates slightly more complex multi-agent systems, which increases inference time.
Nevertheless, MASFly still achieves the best performance among all compared methods, indicating that its dynamic system generation mechanism scales effectively with task complexity and continues to provide substantial accuracy gains without excessive overhead.

\begin{table}[t]
	\centering
	\caption{Per-stage cost breakdown of MASFly.}
	\label{tab:stage_cost}
	\vspace{-4px}
	\resizebox{\columnwidth}{!}{%
		\begin{tabular}{lccccc}
			\toprule
			\textbf{Dataset} & \textbf{Instantiation} & \textbf{Execution} & \textbf{Supervision} & \textbf{Reflection} & \textbf{Total} \\
			\midrule
			TravelPlanner & 4.7\% & 73.5\% & 15.8\% & 6.0\% & \$0.31 \\
			GAIA          & 4.4\% & 78.4\% & 12.4\% & 4.8\% & \$0.07 \\
			\bottomrule
		\end{tabular}%
	}
\end{table}

Table~\ref{tab:stage_cost} further reports the per-stage token cost breakdown of MASFly together with the total API cost per query.
Agent execution dominates the overall cost, accounting for 73.5\% and 78.4\% on TravelPlanner and GAIA, respectively.
Watcher supervision introduces a moderate overhead, whereas system instantiation accounts for less than 5\%.
Reflection is performed only during offline repository construction and does not contribute to held-out test-time inference cost.

\subsection{Impact of Watcher Supervision Frequency}
\label{appendix:watcher}
\begin{figure}[ht]
	\centering
	\vspace{-6px}
	\includegraphics[width=0.48\textwidth]{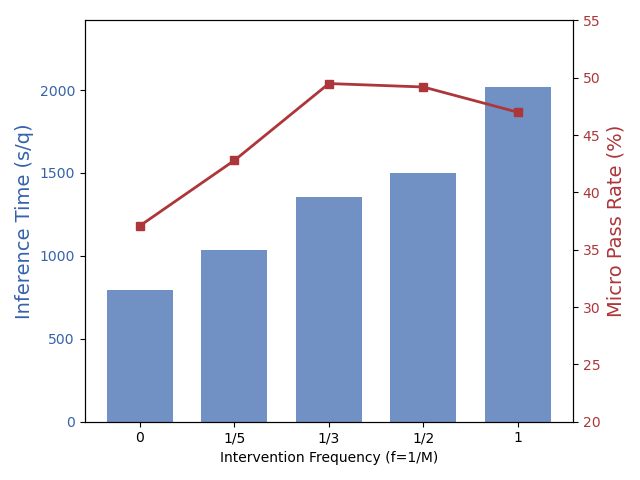}
	\vspace{-14px}
	\caption{The control-efficiency trade-off of supervision module on TravelPlanner. }
	\label{fig:watcher}
	\vspace{-10px}
\end{figure}

To analyze the trade-off between robustness and efficiency in process supervision, we vary the Watcher’s supervision frequency on TravelPlanner, where teams typically consist of 5–6 agents.
Specifically, we set the frequency as
$f=\frac{1}{M}$ where $M$ denotes the number of communication rounds between two supervision events. 
Thus, $f=0$ indicates no supervision and larger $f$ indicates more frequent supervision.
We evaluate both the micro-level hard constraint pass rate and the average execution time per query.

As shown in Figure~\ref{fig:watcher}, moderate intervention substantially enhances robustness by enabling the Watcher to detect abnormal behaviors—such as early-stage communication drift or potential tool misuse—and correct them before they propagate.
However, overly frequent supervision leads to diminishing returns: excessive interventions may introduce unnecessary corrections that disrupt the natural coordination flow among agents, ultimately harming overall stability.
In terms of efficiency, the average execution time increases almost monotonically with supervision frequency, suggesting that, if needed, lowering the intervention frequency can help save computational cost.
Overall, the results indicate that intermediate intervention levels (e.g., setting the interval to half or one-third of the number of agents) offer the best balance between execution robustness and computational efficiency, validating MASFly’s flexible and tunable supervision policy.

\begin{table}[ht]
	\centering
	\caption{Watcher intervention quality on 50 sampled calls from TravelPlanner.}
	\label{tab:watcher}
		\small
	\vspace{-8px}
	\begin{tabular}{lcc}
		\toprule
		& \textbf{No Error} & \textbf{Error} \\
		\midrule
		\textbf{Intervened: Yes} & 6 (FP) & 25 (TP) \\
		\textbf{Intervened: No}  & 16 (TN) & 3 (FN) \\
		\midrule
		\multicolumn{3}{l}{Precision=0.806\quad Recall=0.893\quad F1=0.847} \\
		\bottomrule
	\end{tabular}
\end{table}

\subsection{Watcher Intervention Quality Analysis}

To quantitatively evaluate the reliability of the Watcher, we manually inspect 50 calls of Watcher sampled from TravelPlanner. 
For each call, we annotate (1) whether the underlying system execution truly contained an error, and (2) whether the Watcher issued a proper replacement intervention.

Table~\ref{tab:watcher} summarizes the results. 
As shown, the relatively high recall indicates that most genuine execution errors are successfully detected and corrected. Meanwhile, the precision above 0.8 suggests that erroneous or unnecessary interventions are limited.
We observe that false positives (6 cases) mainly occur when the Watcher leverages its own world knowledge to question intermediate results that are actually consistent within the sandbox environment. 
This discrepancy arises because, by design, the Watcher does not directly observe the full sandbox tool outputs accessed by worker agents. 

\paragraph{False-positive example.}
In one TravelPlanner case, the accommodation returned by the search tool was named ``Room in Down town Brooklyn Parkslop'' for San Francisco.
The Watcher associated ``Brooklyn'' with New York based on its world knowledge and incorrectly inferred a city mismatch, triggering unnecessary replacement of the Accommodation Planner.
The worker agent had in fact obtained the accommodation from the sandbox search tool, but the Watcher did not observe the complete tool-return evidence.
This example illustrates a limitation of supervision based on partial execution context and motivates exposing more structured tool evidence or introducing explicit evidence-consistency checks before agent replacement.

On the other hand, the majority of true positive interventions (25 cases) can be attributed to the Personalized Experience Pool, which provides relevant historical failure patterns and corrective strategies. 
This supports our design choice that experience-enhanced supervision substantially improves robust error detection and recovery.

Overall, the results suggest that the Watcher is not acting as an arbitrary perturbation mechanism, but rather performs reasonably accurate and experience-enhanced supervision.

\subsection{Effect of Different Watcher Backbones}

\begin{table}[ht]
	\centering
	\caption{Effect of different Watcher backbones.}
	\vspace{-4px}
	\setlength\tabcolsep{6pt}
	\small
	\begin{tabular}{lcccc}
		\toprule
		\multirow{2}{*}{Watcher} & \multicolumn{2}{c}{TravelPlanner} & \multicolumn{2}{c}{GAIA}\\
		\cmidrule(lr){2-3} \cmidrule(lr){4-5} 
		& Common & Hard & Level 1 & Level 2  \\
		\midrule
		None  & 83.9  & 37.1  & 41.4 & 36.8 \\
		Qwen3-8B & 84.9 & 47.4 & 43.3 & 38.7 \\
		Qwen3-235B & 85.3 & 49.5 & 46.3 & 41.5 \\
		\bottomrule
	\end{tabular}
	\label{tab:differ_watcher}
	\vspace{-10px}
\end{table}

We further study whether MASFly can benefit from a lightweight Watcher.
Table~\ref{tab:differ_watcher} compares three settings: removing the Watcher entirely, using a lightweight Qwen3-8B Watcher, and using the default Qwen3-235B Watcher.

The results show that even a lightweight Watcher substantially improves performance over the variant without a Watcher.
This indicates that effective process supervision does not strictly require a very large model: a small Watcher is already able to detect many execution failures and provide useful corrective guidance.
At the same time, the stronger Qwen3-235B Watcher consistently achieves the best results, suggesting that more capable Watchers can further improve intervention quality and recovery effectiveness.
Overall, these results show that the benefit of the Watcher is robust across model scales.

\subsection{Robustness under Prompt Attacks}

\begin{figure}[ht]
	\centering
	\vspace{-8px}
	\includegraphics[width=0.47\textwidth]{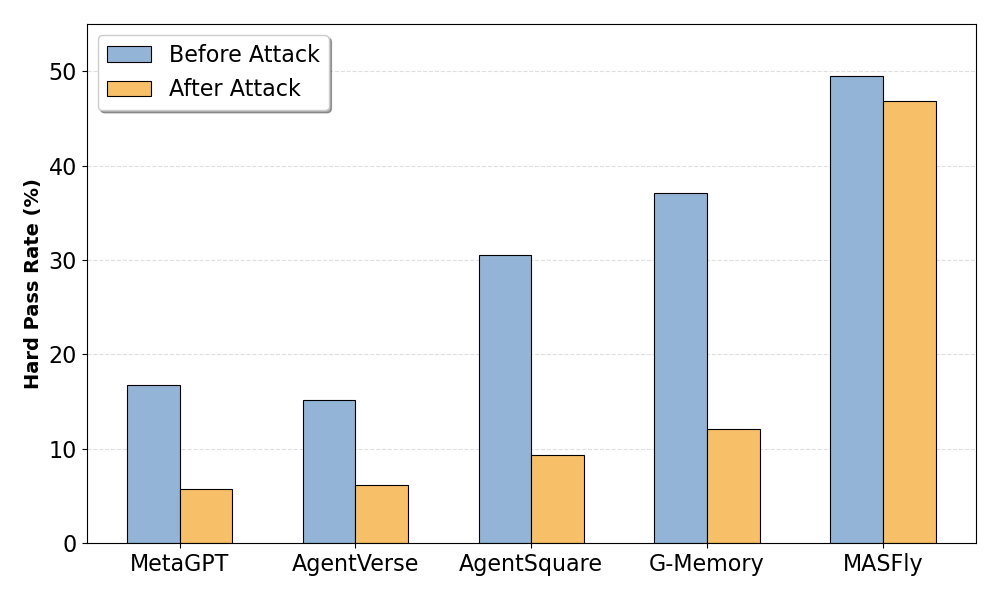}
	\vspace{-10px}
	\caption{Performance before and after prompt injection attack on TravelPlanner.}
	\label{fig:attack-2}
	\vspace{-10px}
\end{figure}

We further evaluate robustness under a more realistic prompt injection attack on TravelPlanner.
Specifically, we inject an adversarial instruction into the restaurant-planning agent, forcing it to avoid tool use and instead generate restaurant plans purely based on its own knowledge.
This setting mimics a common real-world failure mode of LLM agents: bypassing environment interaction and hallucinating plausible but sandbox-inconsistent information.
Figure~\ref{fig:attack-2} shows that baseline methods suffer substantial performance drops under this attack.
Because they rely on open-loop execution, the compromised agent tends to output hallucinated information that does not match the sandbox environment, which severely harms downstream planning quality.
In contrast, MASFly exhibits only a minor degradation.
As Watcher can identify abnormal tool-usage behavior, then trigger supervision and correction during execution.
These results further demonstrate that MASFly is robust not only to explicit failures, but also to realistic prompt-induced hallucinations arising from tool bypass.

\subsection{Evaluation on LiveCodeBench}

\begin{figure}[ht]
	\centering
	\vspace{-8px}
	\includegraphics[width=0.47\textwidth]{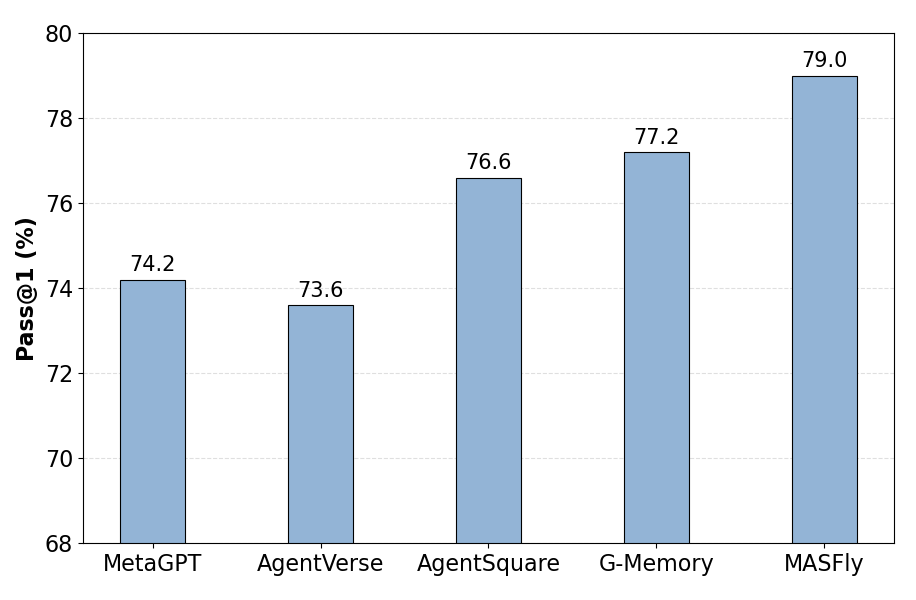}
	\vspace{-10px}
	\caption{Performance comparison on LiveCodeBench-v5 using Qwen3-235B.}
	\label{fig:livecode}
	\vspace{-10px}
\end{figure}

To further evaluate MASFly on a more challenging coding benchmark, we add experiments on \textit{LiveCodeBench}~\cite{jain2024livecodebench} benchmark, which is more competitive than HumanEval and MBPP Pro for modern LLM-based systems.

Figure~\ref{fig:livecode} reports the performance of different methods on LiveCodeBench.
MASFly achieves the best overall result, outperforming all compared representative baselines, including manually designed pipelines (MetaGPT), automated multi-agent frameworks (AgentVerse and AgentSquare), and the memory-augmented baseline G-Memory.
These results further strengthen our conclusions on coding tasks, showing that the advantage of MASFly is not limited to relatively easier benchmarks, but also extends to a more challenging and discriminative coding setting.

\subsection{Effect of Hybrid Retriever}

\begin{table}[ht]
	\centering
	\small
	\caption{Comparison of different retrievers in MASFly.}
	\vspace{-8px}
	\setlength\tabcolsep{4.5pt}
\begin{tabular}{lcccc}
	\toprule
	\multirow{2}{*}{Retrieval Method} & \multicolumn{2}{c}{TravelPlanner} & \multicolumn{2}{c}{GAIA}\\
	\cmidrule(lr){2-3} \cmidrule(lr){4-5} 
	& Common & Hard & Level 1 & Level 2  \\
	\midrule
	QuerySim ($\lambda=1$) & 84.4 & 44.0 &  43.9 & 39.6 \\
	Hybrid ($\lambda=0.8$) & 84.6 & 47.8 & 45.7 & \textbf{41.5} \\
	Hybrid ($\lambda=0.5$) & \textbf{85.3} & 49.1 & 45.7 & \textbf{41.5} \\
	Hybrid ($\lambda=0.3$) & \textbf{85.3} &\textbf{49.5} &\textbf{46.3} & \textbf{41.5} \\
	NeedSim ($\lambda=0$)  & 85.1  &  47.9  &  45.1  & \textbf{41.5} \\
	\bottomrule
\end{tabular}
\label{tab:hybrid}
\vspace{-6px}
\end{table}

Table \ref{tab:hybrid} compares three retrieval strategies: (i) QuerySim, which retrieves SOPs solely based on input query similarity; (ii) NeedSim, which relies only on internal need-level similarity; and (iii) our Hybrid retriever, which integrates both signals.
Compared with using either signal alone, the Hybrid approach yields consistent gains, especially on more challenging cases, demonstrating that external query cues and internal structural needs provide complementary information.
The results also show that performance remained relatively stable across different $\lambda$ (weighting coefficient in the hybrid retriever) values, indicating that MASFly is not highly sensitive to this hyperparameter.
Overall, combining both signals allows the system to retrieve more relevant historical experience from SOP repository and PEP, and thus support more effective multi-agent organization and execution.

\begin{table}[ht]
	\centering
	\small
	\caption{Sensitivity to the number of retrieved SOPs ($K$) in MASFly.}
	\vspace{-4px}
	\begin{tabular}{lcccc}
		\toprule
		\multirow{2}{*}{$K$} & \multicolumn{2}{c}{GAIA} & \multirow{2}{*}{HumanEval} & \multirow{2}{*}{MBPP} \\
		\cmidrule(lr){2-3}
		& Level 1 & Level 2 &  &  \\
		\midrule
		1 & 45.1 & 39.6 & \textbf{98.7} & \textbf{93.3} \\
		2 & \textbf{46.3} & \textbf{41.5} & \textbf{98.7} & \textbf{93.3} \\
		3 & 45.7 & \textbf{41.5} & 98.3 & 92.9 \\
		\bottomrule
	\end{tabular}
	\label{tab:number_of_sop}
	\vspace{-6px}
\end{table}

\subsection{Sensitivity to the Number of Retrieved SOPs}

We further analyze how the performance of MASFly is affected by the number of retrieved SOPs ($K$).
Table~\ref{tab:number_of_sop} reports results when retrieving 1–3 candidate SOPs across three benchmarks.
Overall, MASFly demonstrates stable performance across different values of $K$.
Retrieving two SOPs achieves the best or near-best results on most datasets.
Increasing $K$ beyond two brings limited additional benefit and may slightly reduce accuracy on some tasks, likely because additional retrieved SOPs introduce redundant or less relevant organizational patterns.
These results indicate that MASFly is not highly sensitive to the choice of $K$ within a reasonable range,
and suggests that MASFly benefits from a small set of high-quality organizational priors rather than a large set of retrieved candidates.

\section{Case Study}

We present two SOP examples in Figure~\ref{fig:SOP1} and~\ref{fig:SOP2}, drawn from the GAIA benchmark and a coding task, respectively.
Each SOP specifies the team composition, communication structure, and the detailed configuration of each agent, including responsibilities, instructions, and accessible tools.
These examples demonstrate how SOPs encode structured and interpretable collaboration patterns that can be instantiated for new tasks, where instantiation is mainly achieved by customizing agent instructions and, when necessary, dynamically adjusting other system configurations according to the specific query.

\definecolor{Periwinkle}{RGB}{255, 235, 205}
\definecolor{LightCoral}{RGB}{255, 204, 204}
\definecolor{011_1}{RGB}{184, 183, 163}
\definecolor{011_2}{RGB}{221, 190, 169}
\definecolor{011_3}{RGB}{253, 232, 213}

\newpage
\begin{figure*}[t]
\begin{tcolorbox}[notitle, sharp corners, breakable, 
	colframe=Periwinkle, colback=white, 
	boxrule=3pt, boxsep=0.5pt, enhanced, 
	shadow={3pt}{-3pt}{0pt}{opacity=0.3}]
	\footnotesize
	{\fontfamily{pcr}\selectfont
		\spaceskip=0pt plus 0pt minus 0pt
\begin{lstlisting}[    breaklines=true,
	showstringspaces=false,
	basicstyle=\linespread{2}, % 增加行高
	columns=fixed,          % 保持等宽对齐
	]
{
"team": ["Planner", "WebSearcher", "Summarizer"],
"Communication Structure": "1. User -> Planner;\n2. Planner -> WebSearcher;\n3. WebSearcher -> Summarizer;\n4. Summarizer -> End.\n\n
**Description:**\n1. First, the Planner analyzes the user's query and creates a strategic search plan.\n2. Second, the WebSearcher receives the plan and executes it by gathering the necessary information from the web.\n3. Finally, the Summarizer receives the data from the WebSearcher and synthesizes it into the final answer for the user.",        
"Agent Specifications": [
	{
	"name": "Planner",
	"responsibility": "You are a strategic Planner. Your primary role is to analyze the user's question and formulate a clear, step-by-step plan for the web search. ",
	"instruction": "Break down complex questions into smaller, logical sub-tasks. For each step, define the specific information that need to be searched to effectively answer the user's query. The output of your plan will be a clear set of instructions for the WebSearcher to follow.",
	"tools": []
	},
	{
	"name": "WebSearcher",
	"responsibility": "You are an expert Web Searcher. Your mission is to execute the search according to the search strategy provided by the Planner. Your goal is to gather accurate and relevant information from the web to answer the user's question.",
	"instruction": "1. Strategize Your Search:\nCarefully analyze the request from the PlanAgent. Identify the core pieces of information you need to find. Based on the task, formulate a step-by-step search strategy. Do not try to solve everything with a single query when the query is complex.\n2. Execute the Search:\nUse the GOOGLE Search tool to execute the search. Make sure to follow the usage format of given tool. Each query should focus on key terms from the question. Format the query content as a comma-separated list. \n3. Analyze and Iterate: Review the search results provided in the `Observation`.If the information is sufficient: Synthesize the key findings and prepare to pass them to the AnswerAgent. If the information is insufficient or irrelevant: Return to Step 1. Revise your strategy and create a new query. If you receive an error in the `Feedback`: Analyze the feedback message to understand the issue and adjust your action accordingly.",
	"tools": ["GOOGLE Search"]
	},
	{
	"name": "Summarizer",
	"responsibility": "You are the final Answer Agent. Your core mission is to generate a clear, accurate, and comprehensive answer for the user.",
	"instruction": "Synthesize all the information and data provided by the WebSearcher. Ensure the final response directly addresses all parts of the user's original question, is well-organized, and easy to understand.",
	"tools": []
	}
	]
}
\end{lstlisting}
	}
\end{tcolorbox}
\caption{An SOP example for GAIA benchmark.}
\label{fig:SOP1}
\end{figure*}
\definecolor{Periwinkle}{RGB}{204, 204, 255}
\begin{figure*}[t]
\begin{tcolorbox}[notitle, sharp corners, breakable, 
	colframe=Periwinkle, colback=white, 
	boxrule=3pt, boxsep=0.5pt, enhanced, 
	shadow={3pt}{-3pt}{0pt}{opacity=0.3},
	title={},]
	\footnotesize
	{\fontfamily{pcr}\selectfont
		\spaceskip=0pt plus 0pt minus 0pt 
		
\begin{lstlisting}[breaklines=true,showstringspaces=false,basicstyle=\linespread{3}, % 增加行高
	columns=fixed,          % 保持等宽对齐
	]
{
"team": ["Programming Expert", "Test Analyst", "AnswerAgent"],
"Communication Structure": "1. User -> Programming Expert; 2. Programming Expert -> Test Analyst; 3. Test Analyst -> Programming Expert (if errors) | AnswerAgent (if correct); 4. AnswerAgent -> End.\n\n
**Description:**\nFirst, the Programming Expert implements the solution based on the function signature and docstring. Next, the Test Analyst generates test cases and validates the code. If issues are found, it loops back to the Programming Expert for fixes until all tests pass. Finally, the AnswerAgent delivers the validated final answer.",
"Agent Specifications": [
	{
	"name": "Programming Expert",
	"responsibility": "You are a programming expert. Given a function signature and its docstring by the user, your task is to write the complete and correct code implementation.",
	"instruction": "1. Carefully analysis the user's query function and example case, and write the full code based on the function signature.\n2. Use a Python code block for your response.\n3. Do not change the original function names or input variable types.",
	"tools": []
	},
	{
	"name": "Test Analyst",
	"responsibility": "You are a test analyst. Given a function signature, its docstring, and the candidate solution, your task is to validate the code's correctness.",
	"instruction": "1. Based on the given test example, generate other comprehensive test cases. Do not question the correctness of the given test example.\n2. Use the terminal tool to execute the tests against the code.\n3. If any test fails or errors occur, return the task to the Programming Expert with detailed feedback. If all tests pass, submit the validated solution to the AnswerAgent.",
	"tools": ["bash"]
	},
	{
	"name": "AnswerAgent",
	"responsibility": "You are an AnswerAgent. Your task is to deliver the final code to the user.",
	"instruction": "Present the final code to the user, ensuring it is complete and correct.",
	"tools": []
	}
	]
}
\end{lstlisting}
}
\end{tcolorbox}
\caption{An SOP example for Coding tasks.}
\label{fig:SOP2}
\end{figure*}

\end{document}